\documentclass[aps,preprint,amsmath,amssymb,groupedaddress,preprint]{revtex4-2}
\usepackage{amsmath}
\usepackage{graphicx}
\usepackage{color}
\usepackage{color}
\usepackage{graphicx}
\usepackage{natbib}
\usepackage{epsfig}
\usepackage{setspace}
\usepackage{amsmath,amssymb}
\usepackage{verbatim}
\usepackage{bibentry}
\usepackage{xcolor}
\usepackage{bm}
\usepackage{lipsum}

\begin{document}
	\title{Enhanced thermopower in two-dimensional ruthenium dichalcogenides $RuX_2$ (X = S, Se): a first-principles study}
	
	\author{Parbati Senapati}
	\affiliation{Department of Physics,
		Indian Institute of Technology Patna, Bihta, Bihar, 801106, India}
		\author{Ajay Kumar}
		\affiliation{Department of Physics,
			Indian Institute of Technology Patna, Bihta, Bihar, 801106, India}
	\author{Prakash Parida}\email{pparida@iitp.ac.in}
	\affiliation{Department of Physics,
		Indian Institute of Technology Patna, Bihta, Bihar, 801106, India}

\begin{abstract}{Transition metal dichalcogenides (TMDs) have garnered attention for their potential in thermoelectric applications due to their unique electronic properties and tunable bandgaps. In this study, we systematically explore the electronic and thermoelectric properties of $T^{\prime}-RuX_2$ (X = S, Se) using first-principles calculations and semi-classical Boltzmann transport equations. Our findings confirm that $T^{\prime}-RuX_2$ is energetically and mechanically stable, with high thermopower values such that $T^{\prime}-RuS_2$ exhibits a Seebeck coefficient of $2685~\mu V/K$ for hole doping and $2585~\mu V/K$ for electron doping, while  $T^{\prime}-RuSe_2$ shows values of $1515~\mu V/K$ and $1533~\mu V/K$ for hole and electron doping, respectively. Both materials exhibit reasonable power factors and $ZT$ values, with p-type $T^{\prime}-RuS_2$ and $T^{\prime}-RuSe_2$ achieving maximum ZT values of 0.85 and 0.87, respectively, at 1200~K along the y-direction. These results highlight $T^{\prime}$-$RuS_2$ and $T^{\prime}$-$RuSe_2$ as promising candidates for high-temperature TMD-based thermoelectric devices.}
\end{abstract}

\maketitle
\section{Introduction}
Recently, two-dimensional (2D) thermoelectric materials have attracted considerable interest due to their ability to transform heat into electrical energy \cite{li2020recent}. Nonetheless, the development and identification of materials with superior thermoelectric performance, which is measured by the thermoelectric figure of merit ($ZT$) \cite{snyder2017figure}, remains a formidable challenge \cite{liu2015current}. The $ZT$ value is defined as $ZT = \frac{S^{2} \sigma T}{k}$, where $S$ is the Seebeck coefficient, $\sigma$ is electrical conductivity, and $k$ is thermal conductivity. The thermal conductivity ($k$) can be further broken down into electronic ($k_e$) and phonon ($k_{ph}$) contributions, with $k_{ph}$ arising from lattice vibrations and $k_e$ from electron movement \cite{senapati2024charge}. The field of materials research continues to be captivated by 2D materials because of their unique properties, making them ideal for applications in electronics, thermoelectrics, and optoelectronics \cite{glavin2020emerging,yang2017gas,kim2015materials,gupta2023ultralow}. 

In recent decades, advances in theoretical modeling and experimental synthesis techniques have facilitated the exploration of various 2D materials, both computationally and practically \cite{kumar2024theoretical,senapati2023thermoelectric,liu2019recent}. Examples of such materials include black phosphorus \cite{li2014black}, transition-metal dichalcogenides (TMDs) \cite{feng2020magnetic,feng2020layer,villaos2019thickness,chen2020correlating}, Janus TMDs \cite{tao2020thermoelectric,zhang2020recent,maghirang2019predicting}, group IVA-VA compounds \cite{chang20183d}, MXenes \cite{xin2020mxenes,huang2020large}, and Zintl compounds \cite{peng2018crystal,feng2022prediction}, all of which demonstrate exceptional properties. Consequently, they have found successful applications in electronic, optoelectronic, and thermoelectric devices. 

Several techniques, including epitaxial growth \cite{yuhara2018large}, micromechanical cleavage \cite{xiao2018topochemical}, chemical vapor deposition \cite{xiao2018topochemical}, and  mechanical exfoliation \cite{novoselov2004electric}, have been employed to experimentally synthesize these materials. Notably, 2D TMDs exhibit a combination of high electrical conductivity and low thermal conductivity, making them strong candidates for thermoelectric devices \cite{zhang2017thermoelectric,purwitasari2022high}. This potential has been confirmed through the high thermoelectric performance exhibited by Mo- and W-based TMDs \cite{huang2014theoretical,chen2015thermoelectric}.

In particular, TMDs have a general stoichiometric formula of $MX_2$ because each TMD layer has a central sublayer of transition metal atoms ($M$) positioned between two sublayers of chalcogen atoms ($X = \text{S, Se}$) \cite{kolobov2016two}. These materials have attracted considerable research attention because of their distinctive physical and electronic characteristics. One of the most exciting aspects of TMDs lies in their exfoliation potential \cite{shi2015synthesis}. By using this technique, individual atomic layers (monolayers) can be extracted from bulk TMD crystals. The ability to create high-quality monolayers positions TMDs as prime candidates for developing a new generation of nanoelectronic devices. These devices are expected to be high-performing, energy-efficient, flexible, and tunable. Potential applications include 2D field-effect transistors, light-emitting diodes, biosensors, and phototransistors. The promise of TMDs extends beyond these examples, with reviews highlighting their potential role in emerging technologies like the Internet of Things (IoT) and 5G networks \cite{ajayan2016two}.

Comprehensive theoretical and experimental studies have been carried out on a variety of TMDs. Purwitasari et al. investigated the thermoelectric performance of 2D technetium dichalcogenides (\(TcX_2\), where \(X = S\), \(Se\), or \(Te\)). Their results offer vital insights into the structural stability, strong electronic properties, and good thermoelectric characteristics of \(TcX_2\), underscoring its potential for optoelectronic and thermoelectric applications \cite{purwitasari2022high}. Huang et al. reported that the n-type \(1T\)-MoS\(_2\) and \(2T\)-WSe\(_2\) materials exhibit a favorable thermoelectric figure of merit, highlighting the potential of these TMDs as promising thermoelectric materials \cite{huang2014theoretical}. Yumnam et al. investigated the thermoelectric properties of \(MX_2\) (\(M = Zr, Hf; X = S, Se\)) through density functional theory and Boltzmann transport calculations. They found that n-type doped \(HfSe_2\) exhibits high efficiency for high-temperature thermoelectric applications, achieving a \(ZT_{max} > 1\) at \(1300 \, K\) \cite{yumnam2015high}. Ding et al. conducted an in-depth investigation into the transport properties of $SnSe_2$ using a comprehensive ab-initio study, revealing significant thermopower and a high $ZT$ value in n-type $SnSe_2$ \cite{ding2017transport}. Zhou et al. explored the thermoelectric performance of typical 2D-TMDs, such as $MX_2$ (where $M = Mo, W, Ti; X = S, Se$) and presented relevant theoretical and experimental results. The study highlighted the experimental challenges encountered when investigating the TE properties of these TMDs and emphasized the need for ongoing theoretical and experimental research to advance this field \cite{zhou2022recent}.

However, ruthenium-based TMDs ($RuX_2$, where $X = S, Se$) present a unique opportunity for further exploration due to their limited study. The bulk (3D) structure of the ruthenium-based dichalcogenides has been extensively studied both theoretically and experimentally, showing that these structures are stable with tunable semiconducting electronic properties with a bandgap of $1.3~eV$ \cite{sai2021study,bichsel1984growth,zhao2021hexagonal}. However, our focus is on their 2D counterparts. In 3D materials, electrons and holes can move freely in all directions, resulting in a continuous range of energy levels. When these materials reduced to 2D, the carriers are confined in the third dimension, leading to quantized energy levels. This confinement raises the conduction band and lowers the valence band, effectively widening the bandgap. The confinement of phonon motion to two dimensions also restricts the available phonon modes and reduces their group velocities, resulting in lower thermal conductivity. This increased significant bandgap and lower thermal conductivity in 2D materials makes them more suitable for thermoelectric applications compared to their bulk counterparts. Interestingly, Ersan et al. have demonstrated that 2D-$RuS_2$ (and potentially $RuSe_2$) exhibit unique behavior. Unlike most transition metal dichalcogenides (TMDs), where hexagonal (\(2H\)) and octahedral (\(1T\)) phases are typically stable, these phases are unstable in $RuS_2$ and $RuSe_2$. Instead, the only stable phase for these materials is the \(T'\) (distorted) phase, which is characterized by an orthorhombic unit cell containing six atoms per cell (two $Ru$ and four $S$).
They have been confirmed through first-principles calculations based on DFT that the $T^\prime$ phase is dynamically, mechanically, and thermally stable \cite{ersan2016stable}. Ruthenium-based TMDs possess a wide indirect bandgap. A valence band positioned near the Fermi level, making them promising candidates for thermoelectric applications. Despite these favorable properties, a comprehensive literature review uncovered no prior reports on the thermoelectric characteristics of $T^\prime$ phase of $RuX_2$ TMDs. Hence, investigating the thermoelectric properties of these $RuX_2$ TMDs presents a significant opportunity to explore their potential in this field.

In this paper, we investigate the structural properties, mechanical properties, electronic properties, phonon thermal conductivity, and thermoelectric properties of the 2D TMDs $RuX_2$ ($X$ = $S$, $Se$) using first principle approach. Additionally, we explore various thermoelectric transport parameters, including the electrical conductivity, thermopower,  electronic thermal conductivity, and power factor ($S^{2}\sigma$), employing semi-classical Boltzmann transport equations (BTEs). Understanding phonon transport in low-dimensional systems is crucial for optimizing their thermoelectric properties through nanostructuring and engineering. Then, we calculate the total thermoelectric figure of merit by combining phonon thermal conductivity and electronic thermal conductivity, which gives us a comprehensive measure of overall efficiency of a thermoelectric device.

\section{Computational methods}
We employ first-principles calculations utilizing a plane-wave basis set within the framework of density functional theory (DFT) \cite{hohenberg1964inhomogeneous}, incorporating the projector-augmented wave (PAW) approach \cite{kresse1999ultrasoft}, as implemented in the Vienna Ab initio Simulation Package (VASP) \cite{kresse1996efficient}. The exchange-correlation interactions are described by the Heyd-Scuseria-Ernzerhof (HSE) hybrid functional \cite{heyd2003hybrid,krukau2006influence}, which combines features of both Hartree-Fock and DFT methods. Several convergence tests were conducted to determine an appropriate k-point grid and energy cut-off, with the results provided in Fig. S1 of the Supplementary Information. For k-point sampling (in Fig. S1(a) and (c)), the first Brillouin zone was tested with grid sizes ranging from \(1 \times 1 \times 1\) to \(18 \times 29 \times 1\), and a k-grid of \(9 \times 14 \times 1\) was chosen to ensure computational robustness. We use a Monkhorst-Pack (MP) grid \cite{monkhorst1976special} to ensure precision in both the geometry optimization and electronic structure calculations.  Similarly, the plane wave energy cut-off was examined over a range of 200 to 800 eV. As shown in Fig. S1 (b) and (d), the total energy converges reliably beyond 450 eV. To maintain a balance between computational efficiency and accuracy, a cut-off value of 500 eV was selected. To avoid interactions between periodic images, we maintain a vacuum space greater than \(15 \, \text{\AA}\). The atomic configurations and lattice parameters are optimized using the conjugate gradient technique, continuing until the forces acting on each atom drop below \(0.001 \, \text{eV}/\text{\AA}\) and the variation in total energy between subsequent steps is reduced to less than \(10^{-8} \, \text{eV}\). For the HSE functional, we use a screening parameter of \(0.2 \, \text{\AA}^{-1}\) with a mixing fraction of \(0.25\) for the Hartree-Fock exchange. Band structure computations are performed by sampling the high-symmetry points \(\Gamma-X-S-Y-\Gamma\) to accurately capture the electronic properties of both the structures.

To evaluate the dynamical stability of both structures, we compute the phonon frequencies along key symmetry paths within the 2D Brillouin Zone. This analysis is performed using the PHONOPY software \cite{togo2015first}, which leverages density functional perturbation theory (DFPT) as implemented in the VASP package. We have also performed supercell convergence tests for phonon dispersion, as illustrated in Fig. S2 of Supplementary Information. Based on these tests, we have selected a \(2 \times 3 \times 1\) supercell in our study, ensuring that all long-wavelength vibrational modes are effectively captured for phonon dispersion calculations. For the ab initio molecular dynamics simulations, we similarly adopt a supercell of \(2 \times 3 \times 1\) alongside a k-point grid of \(5 \times 5 \times 1\). To explore the mechanical characteristics, we apply the generalized form of Hooke's law as follows\cite{kumar2024theoretical,eivari2022mechanical}:
\begin{align}
	\sigma^{ij} &= C^{ij}_{\phantom{ij}kl} \, \epsilon^{kl}; \quad C^{ij}_{\phantom{ij}kl} = \frac{1}{2} \frac{\partial^2 U}{\partial \epsilon_{ij} \partial \epsilon^{kl}} \quad (i, j, k, l = 1, 2, 3)
\end{align}

where \(\sigma^{ij}\) represents the contravariant components of the stress tensor, \(\epsilon_{kl}\) represents the second-rank strain tensor, and \(C^{ij}_{\phantom{ij}kl}\) is the fourth-rank stiffness tensor, characterizing the material's response to deformation. These fourth-order stiffness constants \(C^{ij}_{\phantom{ij}kl}\) are expressed in Voigt notation as \(C_{pq}\) \((p,q=1,2,3,4,5,6)\) \cite{barron1965second}. For clarity, the mapping between \(ij\) and \(p\) is as follows: \(11 \rightarrow 1\), \(22 \rightarrow 2\), \(33 \rightarrow 3\), \(12=21 \rightarrow 4\), \(13=31 \rightarrow 5\), and \(23=32 \rightarrow 6\). Similarly, the mapping between \(kl\) and \(q\) is as follows: \(11 \rightarrow 1\), \(22 \rightarrow 2\), \(33 \rightarrow 3\), \(12=21 \rightarrow 4\), \(13=31 \rightarrow 5\), and \(23=32 \rightarrow 6\). This notation simplifies the representation of the stiffness tensor from a fourth-order tensor to a second-order matrix for ease of analysis in the context of the mechanical properties of materials. The elastic strain energy per unit area can be represented as follows:

\begin{align}
	U(\epsilon_{11},\epsilon_{22}) = \frac{1}{2} C_{11} \epsilon_{11}^{2} + \frac{1}{2} C_{22} \epsilon_{22}^{2} + C_{12} \epsilon_{11} \epsilon_{22} + 2 C_{66} \epsilon_{12}^{2}
\end{align}

In this context, $\epsilon_{11}$ and $\epsilon_{22}$ correspond to the strains along the $x$- and $y$-axes, respectively, with $C_{11}$, $C_{22}$, and $C_{12}$ representing the stiffness constants in Voigt notation.

To assess the lattice transport properties, such as phonon thermal conductivity (\( k_{\text{ph}} \)), we use the Phono3py code \cite{togo2015distributions, chaput2013direct}. For this calculation, we employ a supercell of \(2 \times 3 \times 1\) and a \(q\)-grid of \(9 \times 14 \times 1\) for both structures. The convergence of \( k_{\text{ph}} \) with respect to the \(q\)-grid is illustrated in Fig.~S3 (Supplementary Information). We compute the second-order force constants using the Phonopy package with DFPT, applying symmetric displacements to determine the forces for the dynamical matrices. We consistently maintain the same pseudopotentials and plane-wave basis cutoff across all calculations. Fig. S4 in Supplimentary Information illustrates that a 6~\AA\ cutoff for the atomic neighborhood is sufficient to achieve convergence of the third-order force constants, ensuring the accuracy of phonon thermal conductivity calculations. The third-order anharmonic interatomic force constants are computed using a \(2 \times 3 \times 1\) supercell for both \(T^{\prime}\text{-RuS}_2\) and \(T^{\prime}\text{-RuSe}_2\). This supercell generates 2910 displacement datasets for \(T^{\prime}\text{-RuS}_2\) and \(T^{\prime}\text{-RuSe}_2\), with an atomic displacement magnitude of 0.01~\AA. These force constants are then employed to solve the Boltzmann transport equation using the phono3py code.

The phonon thermal conductivity is determined using the single-mode relaxation time approximation (SMRTA) implemented in the Phono3py code \cite{mizokami2018lattice}. It is mathematically expressed as:
	\begin{equation}
		k_{\lambda}^{\alpha\alpha} = \frac{1}{N V} \sum_{\lambda} C_{\lambda} v_{\lambda}^{\alpha} \otimes v_{\lambda}^{\alpha} \tau_{\lambda},
	\end{equation}
	where $C_{\lambda}$ is the phonon heat capacity, defined as:
	\begin{equation}
		C_{\lambda} = k_{B} \frac{\left( \frac{\hbar \omega_{\lambda}}{k_B T} \right)^2 e^{\frac{\hbar \omega_{\lambda}}{k_B T}}}{\left( e^{\frac{\hbar \omega_{\lambda}}{k_B T}} - 1 \right)^2}.
	\end{equation}
	Here, $v_{\lambda}^{\alpha}$ represents the phonon group velocity along the $\alpha$-direction ($\alpha = x/y/z$), given by $v_{\lambda}^{\alpha} = \frac{\partial \omega_{\lambda}}{\partial q_{\alpha}}$. The term $\lambda$ denotes the phonon mode, characterized by the pair of phonon wave vector $\mathbf{q}$ and branch $j$. The parameter $\tau_{\lambda}$ is the phonon relaxation time (or lifetime), derived from the phonon linewidth $2\Gamma_{\lambda}(\omega_{\lambda})$ associated with the phonon mode $\lambda$ \cite{togo2015distributions}:
	\begin{equation}
		\tau_\lambda = \frac{1}{2 \Gamma_\lambda (\omega_\lambda)}.
	\end{equation}
	The expression for $\Gamma_\lambda (\omega_\lambda)$ is obtained using the Fermi golden rule:
	\begin{align}
		\Gamma_\lambda (\omega_\lambda) = \frac{18 \pi}{\hbar^2} 
		\sum_{\lambda' \lambda''} & \left| \phi_{-\lambda \lambda' \lambda''} \right|^2 
		\bigg\{ \left(n_{\lambda'} + n_{\lambda''} + 1 \right) 
		\delta\left(\omega_\lambda - \omega_{\lambda'} - \omega_{\lambda''} \right) \nonumber \\
		&+ \left(n_{\lambda'} - n_{\lambda''} \right) 
		\big[\delta\left(\omega_\lambda + \omega_{\lambda'} - \omega_{\lambda''} \right) 
		- \delta\left(\omega_\lambda - \omega_{\lambda'} + \omega_{\lambda''} \right)\big]
		\bigg\},
	\end{align}
	where $\phi$ defines the interaction strength between three phonons ($\lambda, \lambda', \lambda''$), and $n_\lambda$ represents the phonon occupation number. The analytical derivation of the phonon linewidth is provided in Section I of the Supplementary Information.

	The parameters $N$, $V$, and $T$ correspond to the total number of $\mathbf{q}$-points in the discretized Brillouin zone, the system volume, and the temperature, respectively. Here, $\omega_{\lambda} = \omega(\mathbf{q}, j)$ is the phonon frequency, $k_B$ is the Boltzmann constant, and $\hbar$ is the reduced Planck constant. While $C_{\lambda}$ is typically calculated within the harmonic approximation (assuming linear atomic interactions), the computation of $\tau_{\lambda}$ accounts for anharmonic effects, incorporating phonon-phonon interactions that influence scattering processes.

	To analyze the phonon mode contribution to the thermal conductivity, a density-of-states-like quantity is defined as:
	\begin{equation}
		k_{\lambda}^{\alpha\alpha} (\omega) = \frac{1}{N} \sum_{\lambda} k_{\lambda}^{\alpha\alpha} \delta(\omega - \omega_{\lambda}),
	\end{equation}
	such that the total thermal conductivity, $k_{\text{ph}}^{\alpha\alpha}$, can be expressed as:
	\begin{equation}
		k_{\text{ph}}^{\alpha\alpha} = \int_0^{\infty} k_{\lambda}^{\alpha\alpha} (\omega) \, d\omega,
	\end{equation}
	where $\frac{1}{N} \sum_{\lambda} \delta(\omega - \omega_{\lambda})$ is the phonon density of states.  Additionally, the mode Gr\"{u}neisen parameters, $\gamma_{\lambda}$, at a $\mathbf{q}$-point for phonon mode $\lambda$ are defined as:
	\begin{equation}
		\gamma_{\lambda} = -\frac{V}{\omega_{\lambda}} \frac{\partial \omega_{\lambda}}{\partial V}.
		\end{equation}

For the electronic transport properties, we apply the semi-classical Boltzmann transport equations using the BoltzTraP code \cite{madsen2006boltztrap}, assuming energy-independent relaxation times and using the rigid band approximation. The following tensor equations describe the thermoelectric properties: electrical conductivity $\sigma^{\alpha\beta}$, conductivity driven by the thermal gradient $v^{\alpha\beta}$, and the electronic thermal conductivity $k^{\alpha\beta}_{e}$ in the $\alpha$ and $\beta$ directions.

\begin{align}
	\sigma^{\alpha\beta}(T;\mu)=\frac{1}{\Omega}\int\sigma^{\alpha\beta}(\epsilon)\left[-\frac{\partial f_{\mu}(T,\epsilon)}{\partial\epsilon}\right]d\epsilon
\end{align}
\begin{align}
	v^{\alpha\beta}(T;\mu)=\frac{1}{eT\Omega}\int\sigma^{\alpha\beta}(\epsilon)(\epsilon-\mu)\left[-\frac{\partial f_{\mu}(T,\epsilon)}{\partial\epsilon}\right]d\epsilon
\end{align}
\begin{align}
	k_{e}^{\alpha\beta}(T;\mu)=\frac{1}{e^{2}T\Omega}\int\sigma^{\alpha\beta}(\epsilon)(\epsilon-\mu)^{2}\left[-\frac{\partial f_{\mu}(T,\epsilon)}{\partial\epsilon}\right]d\epsilon
\end{align}
The Seebeck coefficient ($S^{\alpha\beta}$) can be determined from these tensor quantities,
\begin{align}
	S^{\alpha\beta}=\sum_{\gamma}(\sigma^{-1})^{\alpha\gamma}v^{\beta\gamma}
\end{align}
where \(\mu\), \(\Omega\), \(f\), and \(T\) denote the chemical potential, cell volume, Fermi-Dirac distribution, and absolute temperature, respectively. The term \(\sigma^{\alpha\beta}(\epsilon)\) represents the energy-projected conductivity tensor related to the density of states, which is expressed as follows:

\begin{align}
	\sigma^{\alpha\beta}(\epsilon)=\frac{1}{N}\sum_{i,k}\sigma^{\alpha\beta}(i,k)\delta(\epsilon-\epsilon_{i,k})
\end{align}
Here, $N$ represents the total number of k-points sampled, $\epsilon_{i,k}$ denotes the electron band energies for a given band index $i$, and $\sigma^{\alpha\beta}(i,k)$ refers to the components of the conductivity tensor. The term $\delta(\epsilon - \epsilon_{i,k})$ is the Dirac delta function, which ensures that the energy $\epsilon$ matches the energy $\epsilon_{i,k}$ for the electronic state under consideration. A widely used approach to express this conductivity tensor involves the energy-dependent relaxation time approximation (RTA). Within this framework, the conductivity tensor can be written as:

\begin{align}
	\sigma^{\alpha\beta}(i,k)=e^{2}\tau_{i,k}\mathcal{V}^\alpha(i,k)\mathcal{V}^\beta(i,k)
\end{align}
where \(e\) is the elementary charge of the electron, \(\tau_{i,k}\) represents the relaxation time for the electronic state \((i,k)\), and \(\mathcal{V}^\alpha(i,k)\) and \(\mathcal{V}^\beta(i,k)\) denote the group velocities of the electronic state \((i,k)\), which can be expressed as follows: $\mathcal{V}^\alpha(i,k) = \frac{1}{\hbar} \frac{\partial \epsilon_{i,k}}{\partial k_{\alpha}}, \quad \mathcal{V}^\beta(i,k) = \frac{1}{\hbar} \frac{\partial \epsilon_{i,k}}{\partial k_{\beta}}$, where \(\alpha\) and \(\beta\) are tensor indices.

BoltzTraP integrates both electrical and thermal conductivity by including a relaxation time \((\tau)\) within the framework of the Boltzmann transport equation. This relaxation time is derived from the effective mass and charge carrier mobility, which are estimated using deformation potential (DP) theory. To better understand the electronic transport properties of 2D materials, we apply DP theory, introduced by Bardeen and Shockley \cite{bardeen1950deformation}, to calculate the theoretical carrier mobility \((mob_{2D})\). Furthermore, the effective masses of electrons \((m_e)\) and holes \((m_h)\) are obtained by fitting a quadratic function \((E(k) = pk^2 + qk + r)\) to the curvature of the band edges close to the Fermi level, with this fitting process executed using least squares minimization.
The  expression for $m^*$ is, 
\begin{align}
	m^{*}=\frac{\hbar^2}{\frac{\partial^{2}E}{\partial k^{2}}}
\end{align}
Additionally, the mobility of carriers and relaxation time can be determined utilizing the following relations \cite{wang2016carbon,bruzzone2011ab}:
\begin{align}
	mob_{2D}=\frac{e\hbar^{3}C_{2D}}{k_{B}T|m^{*}|^{2}E_{DP}^{2}}
\end{align}
and,
\begin{align}
	\tau=\frac{|m^{*}|}{e}mob_{2D}
\end{align}
where \(k_{B}\) denote the Boltzmann constant, \(T\) is the temperature, and \(m^{*}\) represents the effective mass of the charge carrier. The elastic modulus \(C_{2D}\) is defined as $C_{2D} = \frac{1}{A_0} \frac{\partial^{2}E}{\partial\chi^{2}}$,
where \(E\) correspond to the total energy under various deformation states, \(A_0\), and \(\chi\) are the equilibrium lattice area, and the applied strain, respectively. $C_{2D}$ is determined by fitting a quadratic function to the energy-strain data and $E_{DP}$ is the DP constant which quantifies how the energy levels of electronic bands in a material shift in response to mechanical strain. The deformation potential constant is defined by the equation:  
\begin{equation}
	E_{DP} = \frac{\Delta E_i}{\chi}
\end{equation}
where \( \Delta E_i \) represents the change in energy of the \( i \)-th electronic band (such as the conduction band minimum or valence band maximum) due to applied strain. \( \chi \) denotes the applied strain, calculated as  
\begin{equation}
	\chi = \frac{\Delta A}{A_0}
\end{equation}
with \(A_0\) being the lattice area and \( \Delta A \) the change in lattice area along the transport direction. By systematically varying the strain and measuring the corresponding band energy shifts, we obtained the deformation potential constants for both electrons (CBM) and holes (VBM), which provide insight into how the band edges respond to structural modifications.

\section{Results and discussions}
\subsection{Atomic structures and structural stability}
\begin{figure} 		
	\centering
	\includegraphics[width=1.0\columnwidth]{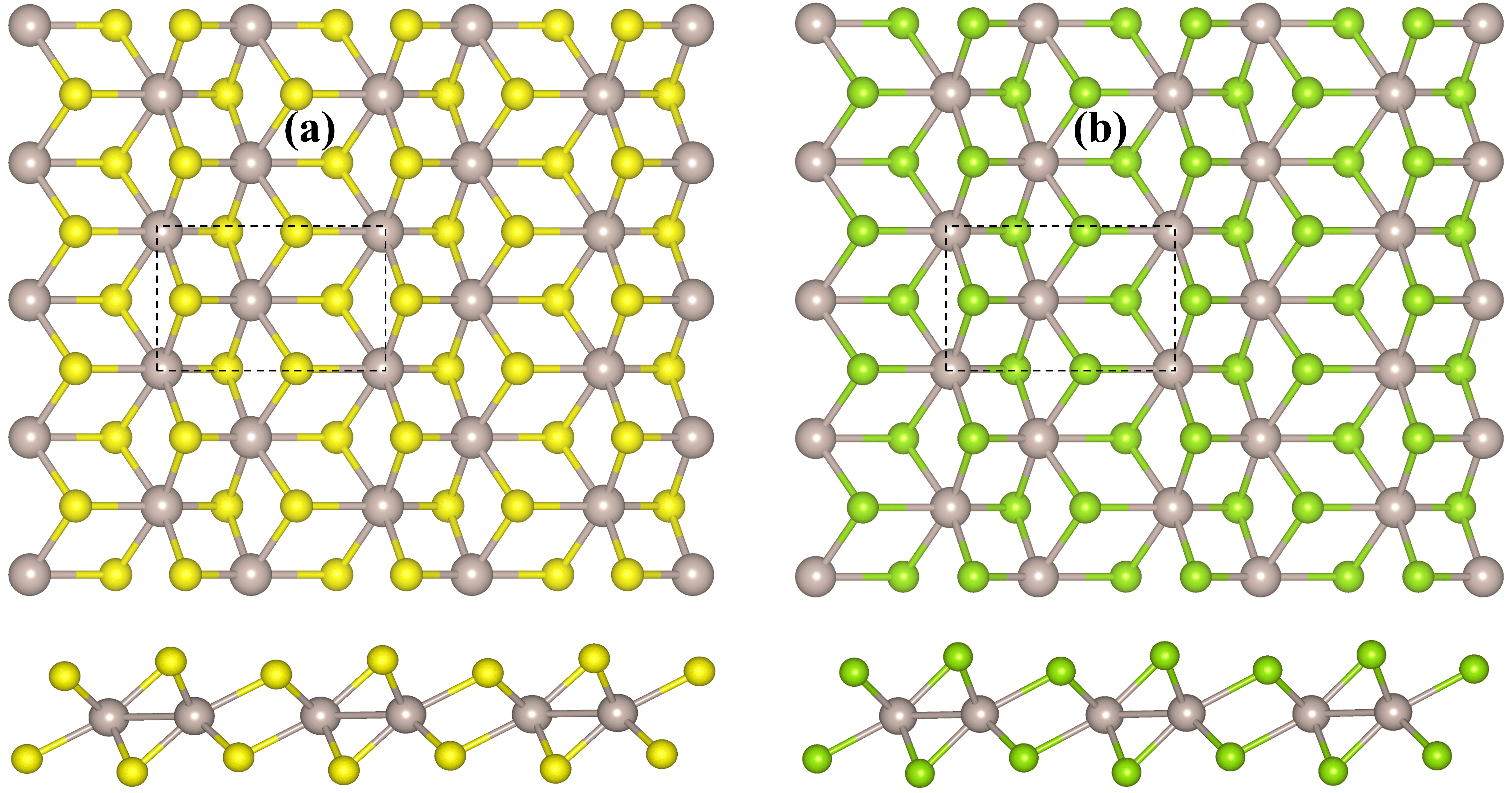}
	\caption{\label{str_plot} (a) and (b) show the top and side views of the optimized structures of $T^{\prime}-RuS_2$ and $T^{\prime}-RuSe_2$, respectively. The grey, yellow, and green spheres represent $Ru$, $S$, and $Se$ atoms, respectively. The black dashed rectangle outlines the primitive cell for each structure.}
\end{figure}

We have investigated the structural and electronic characteristics of $T^{\prime}-RuX_2$ (where $X = S, Se$) utilizing first principles methods. Fig. \ref{str_plot} (a) and (b) depict the top and side views of the atomic structure of $T^{\prime}-RuS_2$ and $T^{\prime}-RuSe_2$, respectively. Both structures comprises a rectangular unit cell and belong to the $P2_{1}/m$ space group. They possess the same atomic structures, differing solely in the types of constituent atomic species. The optimized unit cells have lattice parameters of $a = 5.560~\AA$ and $b = 3.450~\AA$ for $T^{\prime}-RuS_2$, and $a = 5.789~\AA$ and $b = 3.597~\AA$ for $T^{\prime}-RuSe_2$, respectively which align well with previous theoretical values \cite{ersan2016stable}. In both structures, $c$ is set to $15~\AA$ to prevent interaction between the layers. The primitive cell consists of six atoms: two $Ru$ atoms and four $X$ atoms ($X$ comprising $S$ and $Se$).

To assess the structural stabilities, we initially compute the cohesive energy of the $T^{\prime}-RuX_2$ to evaluate their chemical bond strength. The cohesive energy ($E_{coh}$) of the $T^{\prime}-RuX_2$ is calculated as follows:
\begin{align}
	E_{coh}=\frac{E_{RuX_{2}}-(N_{Ru}E_{Ru}+N_{X}E_{X})}{N_{Ru}+N_{X}}
\end{align}
Here, $E_{Ru}$, $E_X$, and $E_{RuX_2}$ denote the energy of an isolated $Ru$ atom, the energy of a single atom of element $X$ (where $X$ can be either S or Se), and the total energy of the $T^{\prime}-RuX_2$ structure, respectively. $N_{Ru}$ and $N_{X}$ is the number of $Ru$ and $X$ atoms in the unit cell.  We calculated the cohesive energies for $T^{\prime}-RuS_2$ and $T^{\prime}-RuSe_2$ to be $-3.352~eV$ and $-2.892~eV$ per atom, respectively. The negative cohesive energies suggest that the $T^{\prime}$ phases are more energetically stable, indicating strong bonds between the constituent atoms.

Cohesive energy is crucial for understanding the internal bonding strength of materials and predicting their behavior under various conditions, such as mechanical stress and thermal vibrations. However, cohesive energy does not directly reflect the stability of a material when compared to its elemental constituents in their most stable phases. To more accurately assess thermodynamic stability, we use formation energy which is mathematically defined as, 
\begin{align}
	E_{form}=\frac{E_{RuX_{2}}-(N_{Ru}\mu_{Ru}+N_{X}\mu_{X})}{N_{Ru}+N_{X}}
\end{align} 
Here, $\mu_{Ru}$ and $\mu_{X}$ represent the chemical potentials of $Ru$ and element $X$, respectively. The chemical potential is determined using the formula $\mu_{Ru/X} = \frac{E_{Ru/X}(\text{bulk})}{m}$, where $m$ denotes the number of $Ru$ or $X$ atoms (with $X$ being either S or Se) present in the bulk structure. The bulk Ru has $P6_3/mmc$ symmetry, while the $X$ atoms ($S, Se$) have $P2/c$ and $P2_1/c$ point-group symmetry, respectively.
Our calculations yield $E_{form}$ values of $-0.389~eV$ for $T^{\prime}-RuS_2$ and $-0.344~eV$ for $T^{\prime}-RuSe_2$. These negative formation energies indicate that both materials are thermodynamically stable relative to their constituent elements and can experimentally be synthesized.

\begin{figure}		
	\centering
	\includegraphics[width=1.0\columnwidth]{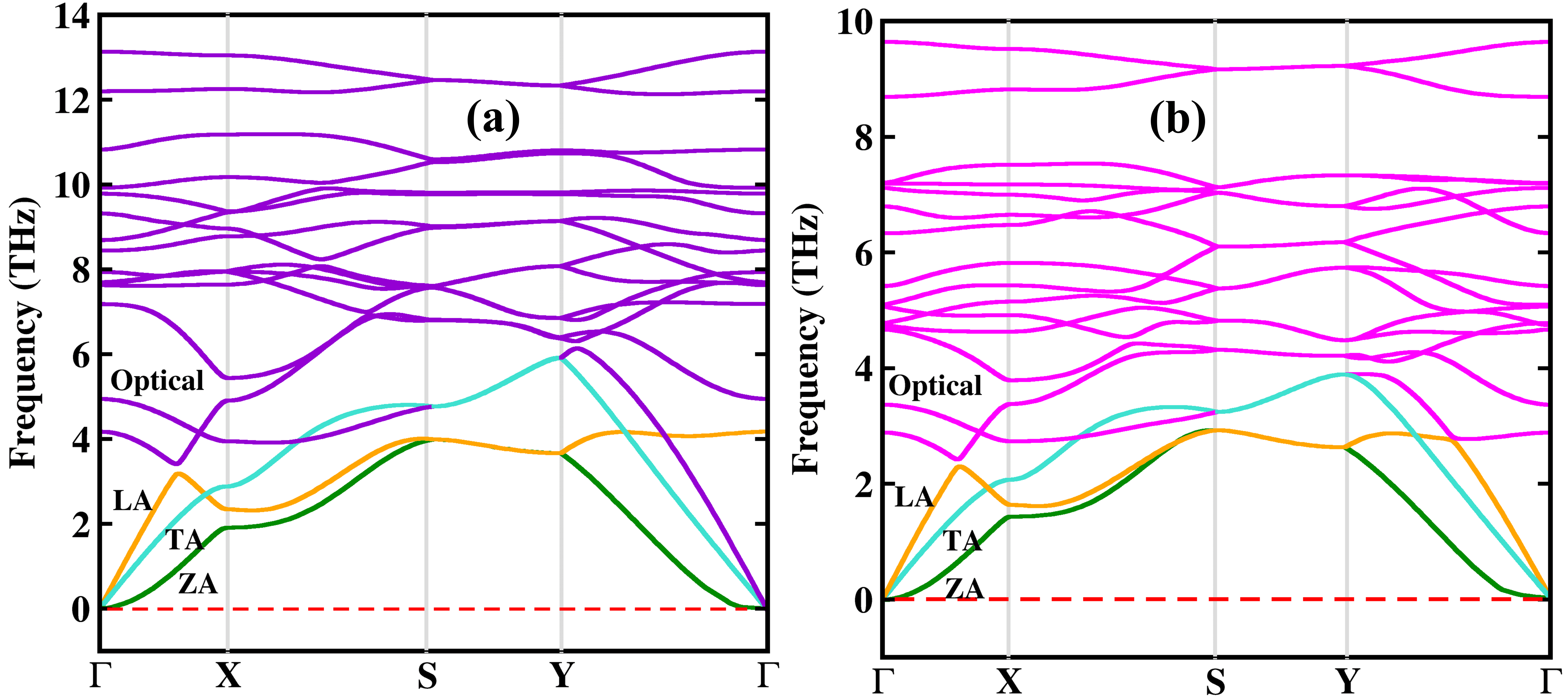}
	\caption{\label{phonon_plot} The phonon dispersion plots of the (a) $T^{\prime}-RuS_2$ and (b) $T^{\prime}-RuSe_2$.}
\end{figure}

To assess the dynamical stability of the $T^{\prime}-RuX_2$ structures, we employed the PHONOPY program to calculate phonon frequencies along the high-symmetry directions within the 2D Brillouin zone. We selected a supercell with dimensions of $2 \times 3 \times 1$ and employed a $9 \times 14 \times 1$ k-point grid to accurately capture long-wavelength vibrational modes, which are essential for conducting stability analysis. The absence of imaginary (negative) phonon frequencies across the entire BZ, as observed in Fig. \ref{phonon_plot} (a) and (b), signifies the dynamical stability of the $T^{\prime}-RuX_2$ structures. The presence of real-valued phonon frequencies throughout the BZ confirms the stability of these materials against phonon-induced distortions. Given that the unit cell of the structure consists of six atoms, this leads to a total of eighteen vibrational modes. Among these, the first three are classified as acoustic modes: the in-plane longitudinal acoustic (LA) mode, the transverse acoustic (TA) mode, and the out-of-plane acoustic (ZA) mode which are labelled in Fig. \ref{phonon_plot} (a) and (b). The remaining fifteen modes are classified as optical modes. 

No gap was detected between the acoustic and optical vibrational spectra in these two structures, indicating strong acoustic-optical scattering. This strong scattering effect enhances phonon-phonon interactions, effectively reducing lattice thermal conductivity, which is desirable for thermoelectric applications. Furthermore, the phonon drag effect, which is governed by phonon dispersion, phonon group velocity, and phonon-electron interactions, exhibits differences between \(T^{\prime}\text{-RuS}_2\) and \(T^{\prime}\text{-RuSe}_2\). In \(T^{\prime}\text{-RuS}_2\), the optical phonon branches extend to a higher frequency ($\approx$14 THz) compared to \(T^{\prime}\text{-RuSe}_2\) ($\approx$10 THz), as shown in Fig. \ref{phonon_plot}. The broader phonon bandwidth in \(T^{\prime}\text{-RuS}_2\) indicates stronger phonon-electron interactions, which enhance the phonon drag effect. Consequently, this results in an increase of the Seebeck coefficient in \(T^{\prime}\text{-RuS}_2\) compared to \(T^{\prime}\text{-RuSe}_2\). \cite{wagner2014phonon}. In \(T^{\prime}\text{-RuS}_2\), low-frequency optical phonons extend below 4 THz (between Y to $\Gamma$ path), enabling interactions with acoustic phonons and contributing to heat transport. These interactions enhance phonon-phonon scattering, significantly influencing thermal conductivity. In contrast, \(T^{\prime}\text{-RuSe}_2\) exhibits weaker contributions from low-frequency optical modes, thereby limiting such interactions and further reducing its thermal conductivity.

\begin{figure}		
	\centering
	\includegraphics[width=0.9\columnwidth]{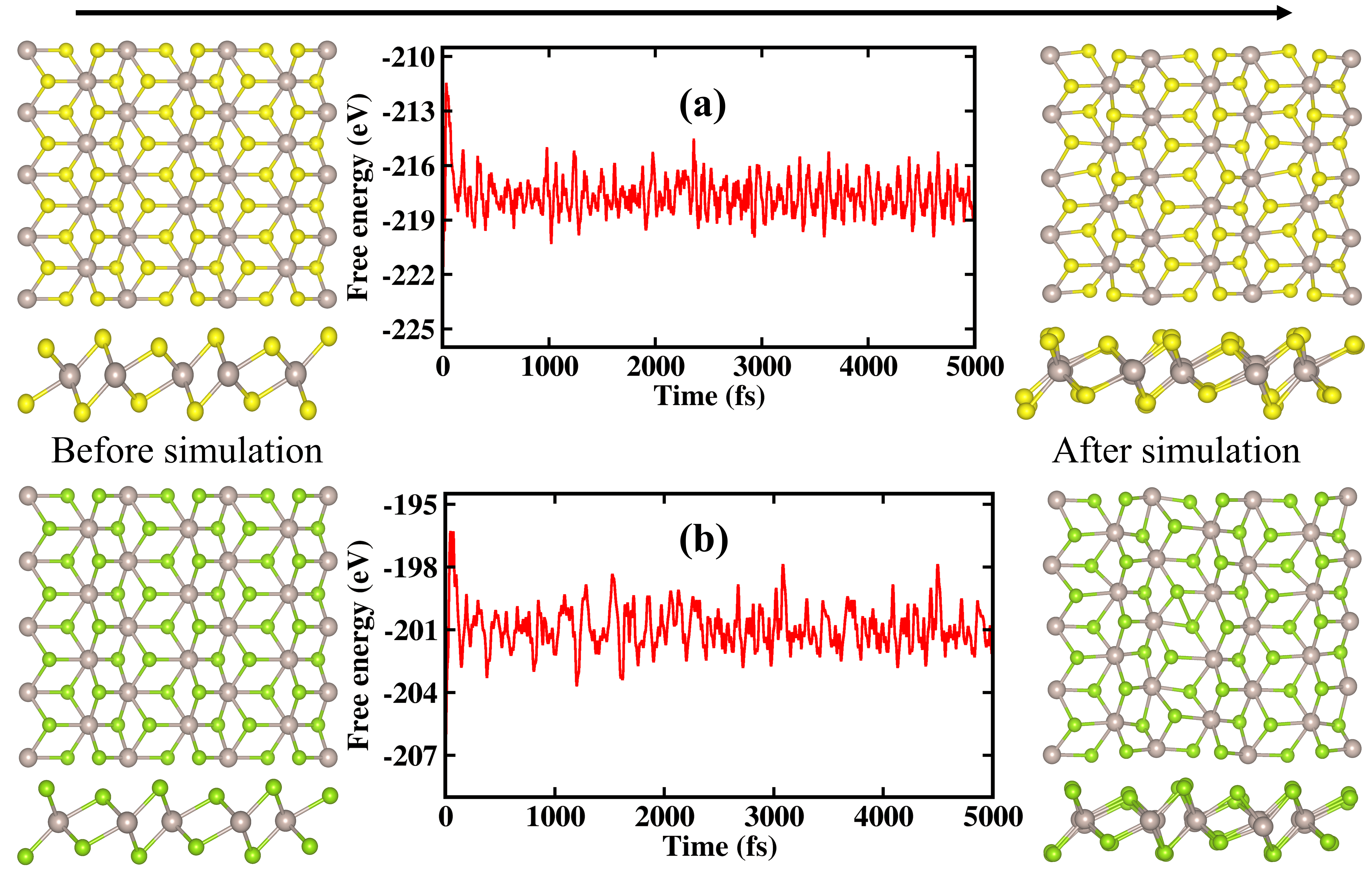}
	\caption{\label{md_plot} Ab initio molecular dynamics calculations of the thermal stability of (a) $T^{\prime}-RuS_2$ and (b) $T^{\prime}-RuSe_2$ structures at  $1200~K$.}
\end{figure}
To investigate the structural stability of $T^{\prime}-RuS_2$ and $T^{\prime}-RuSe_2$ structures at finite temperature, we conducted the ab initio molecular dynamics simulations at $1200~K$ over a simulation period of $5000~fs$, using a $1~fs$ time step. Figure \ref{md_plot} (a) and (b) illustrate the free energy versus time, along with snapshots of both the geometries before and after the simulation. The results show only minor bond stretching without any bond breakage, maintaining the structural stability of $T^{\prime}-RuS_2$ and $T^{\prime}-RuSe_2$ even at high temperatures.  The final geometries exhibit tolerable out-of-plane buckling at $1200~K$, suggesting that thermal vibrations slightly influenced the planar structure. However, the absence of significant bond breaking or structural deformation confirms the stability of these materials at high temperatures. Overall, these findings underscore the thermal stability of both $T^{\prime}-RuS_2$ and $T^{\prime}-RuSe_2$, suggesting their potential suitability for high-temperature applications.

We evaluate the mechanical properties of the $T^{\prime}-RuX_2$ structures by computing their elastic constants, which play a crucial role in determining additional elastic parameters such as Young's modulus ($Y_{2D}$) and Poisson's ratio ($\rho_{2D}$). For 2D structures, four elastic constants need to be evaluated: $C_{11}$, $C_{22}$, $C_{12}$, and $C_{66}$. The elastic constants ($C_{ij}$) of the $T^{\prime}-RuX_2$ structures are listed in Table \ref{tables1}. The calculated elastic constants ($C_{ij}$) reveal that all four values are positive. Additionally, $C_{11}C_{22}> C_{12}^2$ \cite{mouhat2014necessary}. These observations strongly suggest mechanical stability for the $T^{\prime}-RuX_2$ structures. Moreover, Young's modulus and Poisson's ratio for the $T^{\prime}-RuX_2$ structures have been calculated as follows:
$Y_{2D} = \frac{C_{11}^2 - C_{12}^2}{C_{11}}$ and $\rho_{2D} = \frac{C_{12}}{C_{11}}$. The $Y_{2D}$ and $\rho_{2D}$ values for $T^{\prime}-\text{RuS}_2$ are $131.80~\text{N/m}$ and $0.365$, respectively, while for $T^{\prime}-\text{RuSe}_2$ they are $115.88~\text{N/m}$ and $0.360$, respectively, as listed in Table \ref{tables1}. Our findings align closely with previously reported values for various TMDs in the literature, obtained from both theoretical and experimental investigations \cite{sun2021first,singh2018structural,mohanta2020interfacial}. These results indicate that our $T^{\prime}-\text{RuX}_2$ structures possess strong mechanical properties.
 \begin{table} 
 	\centering
 	\caption{Elastic constants ($C_{ij}$), Young's modulus ($Y_{2D}$) and Poisson's ratio ($\rho_{2D}$) for the $T^{\prime}-RuX_2$ structures.}
 	\label{tables1}
 	\begin{tabular}{ccccccc}
 		\hline
 		\hline
 		& {$C_{11}$} & {$C_{12}$} & {$C_{22}$} & {$C_{66}$} & {$Y_{2D}$} & {$\rho_{2D}$}\\
 		& (N/m) & (N/m) & (N/m) & (N/m) & (N/m) & \\ % <-- added & and content for each column
 		\hline
 		{$T^{\prime}-RuS_2$} & 152.06 & 55.50 & 152.06 & 53.40 & 131.80 & 0.365\\ % <--
 		$T^{\prime}-RuSe_2$ & 133.23 & 48.09 & 133.23 & 45.75 & 115.88 & 0.360\\ % <--	
 		\hline
 		\hline
 	\end{tabular}
 \end{table}

\begin{figure}		
	\centering
	\includegraphics[width=1.0\columnwidth]{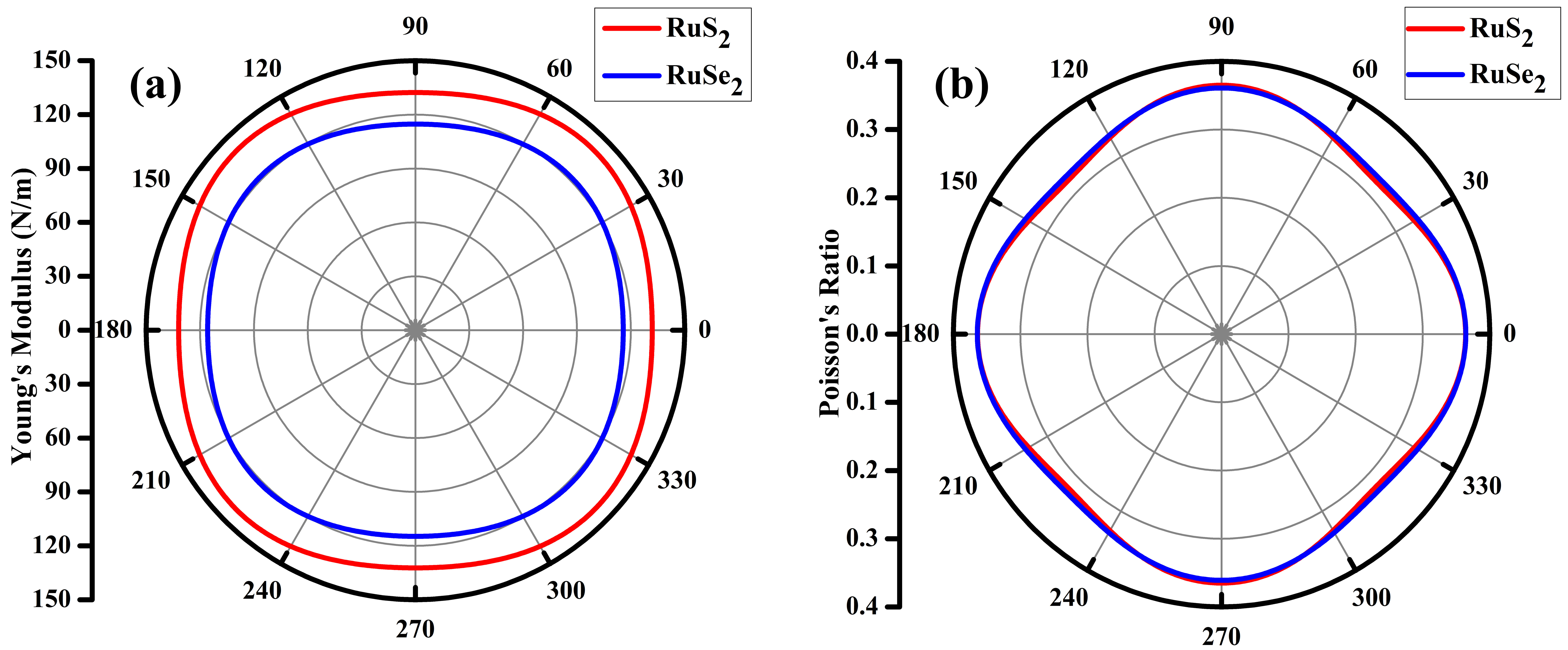}
	\caption{\label{mechanical_plot} Angular variation of (a) Young's modulus and (b) Poisson's ratio of the $T^{\prime}-RuX_2$ structures.}
\end{figure}

Next, we examine the directional dependence of  \(Y_{2D}(\theta)\) and  \(\rho_{2D}(\theta)\) along an arbitrary in-plane direction, specified by the angle $\theta$ with respect to the x-axis. These elastic properties can be expressed in terms of the calculated elastic constants ($C_{ij}$) using the following relationships:
\begin{align}
	Y_{2D} (\theta) = \frac{C_{11}C_{22}-C_{12}^2}{C_{11}\sin^{4}\theta+C_{22}\cos^{4}\theta+(\Lambda-2C_{12})sin^{2}\theta\cos^{2}\theta}
\end{align}
and,
\begin{align}
	\rho_{2D} (\theta) = \frac{(C_{11}+C_{22}-\Lambda) sin^{2}\theta\cos^{2}\theta-C_{12}(\cos^{4}\theta+\sin^{4}\theta)}{C_{11}\sin^{4}\theta+C_{22}\cos^{4}\theta+(\Lambda-2C_{12})sin^{2}\theta\cos^{2}\theta}
\end{align}
where, $\Lambda$=$\frac{C_{11}C_{22}-C_{12}^2}{C_{66}}$.
The detailed analytical derivation of \(Y_{2D}(\theta)\) and  \(\rho_{2D}(\theta)\)  is provided in Section II of the Supplementary Information.

Fig. \ref{mechanical_plot} (a) and (b) illustrates the $\theta$ dependence of Young's modulus ($Y_{2D}$) and Poisson's ratio ($\rho_{2D}$), respectively in $T^{\prime}-RuX_2$. The results indicate that both properties exhibit directional anisotropy, likely due to the inherent lattice anisotropy of $T^{\prime}-RuX_2$. However, the anisotropy of Young's modulus is significantly weaker compared to the anisotropy observed in Poisson's ratio. The difference in Young's modulus in different directions is not significant. For $T^{\prime}-RuS_2$, the minimum and maximum Young's moduli are $132.26~N/m$ (at $\theta=90^\circ$) and $141.07~N/m$ (at  $\theta=44^\circ$), respectively. Similarly, for $T^{\prime}-RuSe_2$, the minimum and maximum Young's moduli are $114.70~N/m$ (at $\theta=90^ \circ$) and $121.51~N/m$ (at $\theta=44^\circ$), respectively. Meanwhile, Poisson's ratio of the $T^{\prime}-RuX_2$ exhibits significant directional anisotropy as shown in Fig. \ref{mechanical_plot}(b). Notably, the Poisson's ratio of $T^{\prime}-RuX_2$ is the highest along the x-axis (at $\theta=0^\circ$).

\subsection{Electronic properties}

\begin{figure}		
	\centering
	\includegraphics[width=1.0\columnwidth]{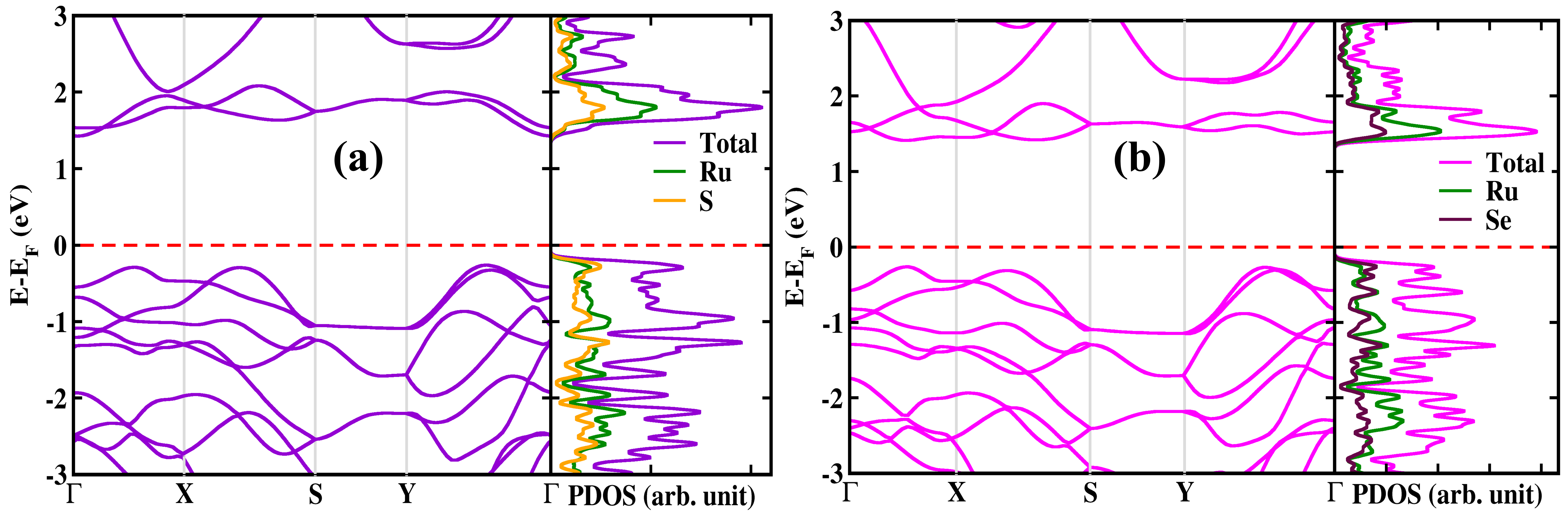}
	\caption{\label{HSE_band_plot} Electronic band spectra and projected density of states (DOS) plots of (a) $T^{\prime}-RuS_2$ and (b) $T^{\prime}-RuSe_2$. The Fermi level (red line) is scaled to zero.}
\end{figure}

The electronic structures provide insights into the carrier transport properties of a material. Thus, we now turn our attention to the electronic structures of the $T^{\prime}-RuX_2$. We used the HSE06 hybrid functional for the band gap determination. In Fig. \ref{HSE_band_plot}, we present the band diagrams of the $T^{\prime}-RuX_2$ structures investigated by using the HSE06 method. The calculations reveal that the $T^{\prime}-RuX_2$ structures are indirect bandgap semiconductors, exhibiting bandgaps of $1.69~eV$ for $T^{\prime}-RuS_2$ and $1.67~eV$ for $T^{\prime}-RuSe_2$. Both the materials show almost identical band dispersion in the vicinity of Fermi energy, differing only in their bandgaps. Thus, $T^{\prime}-RuS_2$ and $T^{\prime}-RuSe_2$ are suitable materials for applications in semiconductor electronics, optoelectronics, and thermoelectrics. For both structures, the VBM are situated along the $Y$ to $\Gamma$ path. The CBM for $T^{\prime}-RuS_2$ is found at the $\Gamma$ point. In the case of $T^{\prime}-RuSe_2$, the VBM is located near the $\Gamma$ point situated between the $\Gamma$ and $X$ points.

To find the contributions to the valence band and conduction band extrema, we show the projected density of states for both structures at the right side of the band structures, as illustrated in Fig. \ref{HSE_band_plot}(a) and (b). In both scenarios, the $Ru$ atom makes a major contribution to the CBM, whereas the VBM is nearly equally contributed by the $Ru$ and the $X$ atom ($S$ and $Se$). Both structures exhibit optimal bandgaps where the valence band and the conduction band edges occur at different k-values, which is suitable for thermoelectric studies. This characteristic is advantageous for thermoelectric applications as it helps maintain a high carrier concentration while minimizing undesirable recombination effects. Moreover, the presence of degenerate states in both the conduction and valence bands enhances the Seebeck coefficient by increasing the number of available states for charge carriers to participate in transport.

\subsection{Phonon thermal conductivity}
\begin{figure}		
	\centering
	\includegraphics[width=1.0\columnwidth]{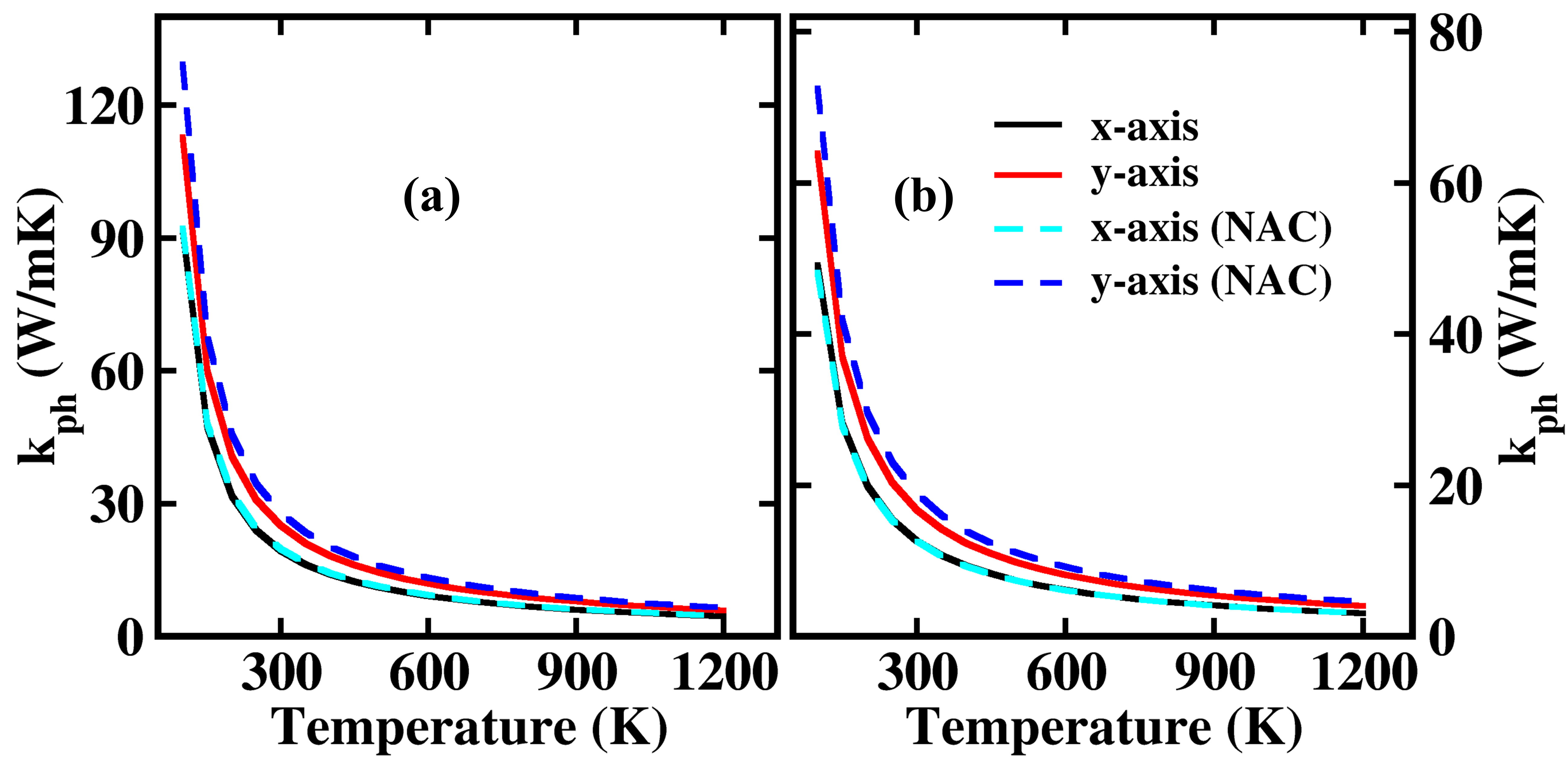}
	\caption{\label{kph_plot} Phonon thermal conductivity as a function of temperature plotted without non-analytical correction (NAC) (solid lines) and with NAC (dashed lines) for (a) $T^{\prime}-RuS_2$ and (b) $T^{\prime}-RuSe_2$.}
\end{figure}

Non-analytical corrections (NACs) are adjustments applied to phonon dispersion relations in polar materials to account for long-range Coulomb interactions between ions. These corrections are essential for accurately capturing the splitting between longitudinal optical (LO) and transverse optical (TO) phonon modes near the Brillouin zone center (\(\Gamma\)-point), commonly known as LO-TO splitting \cite{shafique2020effect}. While NACs play a significant role in highly ionic materials such as oxides and nitrides, their impact is generally minimal in less ionic compounds. In our study of phonon thermal conductivity in \(T^{\prime}\)-RuS\(_2\) and \(T^{\prime}\)-RuSe\(_2\), NACs were included to ensure accuracy. Given that \(T^{\prime}\)-RuS\(_2\) and \(T^{\prime}\)-RuSe\(_2\) exhibit less polarity rather than strong ionicity, the influence of NACs on phonon transport is expected to be negligible. However, to eliminate potential small errors, we incorporated NACs in our calculations. As shown in Fig. \ref{kph_plot} (a) and (b), the phonon thermal conductivity values remain nearly unchanged when comparing results with and without NACs for both \(T^{\prime}\)-RuS\(_2\) and \(T^{\prime}\)-RuSe\(_2\) in x- and y-axis . Since the difference is insignificant, we proceeded with thermoelectric calculations using the phonon thermal conductivity values obtained without NAC inclusion. The temperature-dependent $k_{ph}$ varies as $T^{-1}$ because of phonon-phonon scattering. In $T^{\prime}-RuS_2$ and $T^{\prime}-RuSe_2$, our calculated $k_{ph}$ values (without NAC) at 300~K are 19.30 (25.09)~W/m·K and 12.64 (16.69)~W/m·K in the x- and y-directions, respectively. These low \( k_{ph} \) values are comparable to those of several thermoelectric TMD materials, such as \( SiS_2 \) (\( 15.85~\text{W/mK} \)) \cite{bera2022low} and \( SiSe_2 \) (\( 15.85~\text{W/mK} \)) \cite{bera2022low}. In contrast, they are significantly lower than those of widely studied 2D TMDs like \( MoS_2 \) (\( k_{ph} = 34.5~\text{W/mK} \)) \cite{yan2014thermal} and \( WS_2 \) (\( k_{ph} = 72~\text{W/mK} \)) \cite{bera2019strain}.

Phonon dispersion analysis (in Fig.\ref{phonon_plot}) offers key insights into the origin of the low thermal conductivity ($k_{ph}$) observed in  $T^{\prime}-RuS_2$ and $T^{\prime}-RuSe_2$. In contrast to materials like $MoS_2$ and $WS_2$, which exhibit a well-defined separation between the acoustic and optical branches, the phonon dispersion curves for  $T^{\prime}-RuS_2$ and $T^{\prime}-RuSe_2$ lack such a distinction. The lack of a gap results in significant coupling between the acoustic and optical branches, which is evidenced by the continuous dispersion observed throughout the Brillouin zone. Such coupling is known to enhance phonon-phonon scattering, thereby hindering the propagation of phonons. 

\begin{figure}		
	\centering
	\includegraphics[width=1.0\columnwidth]{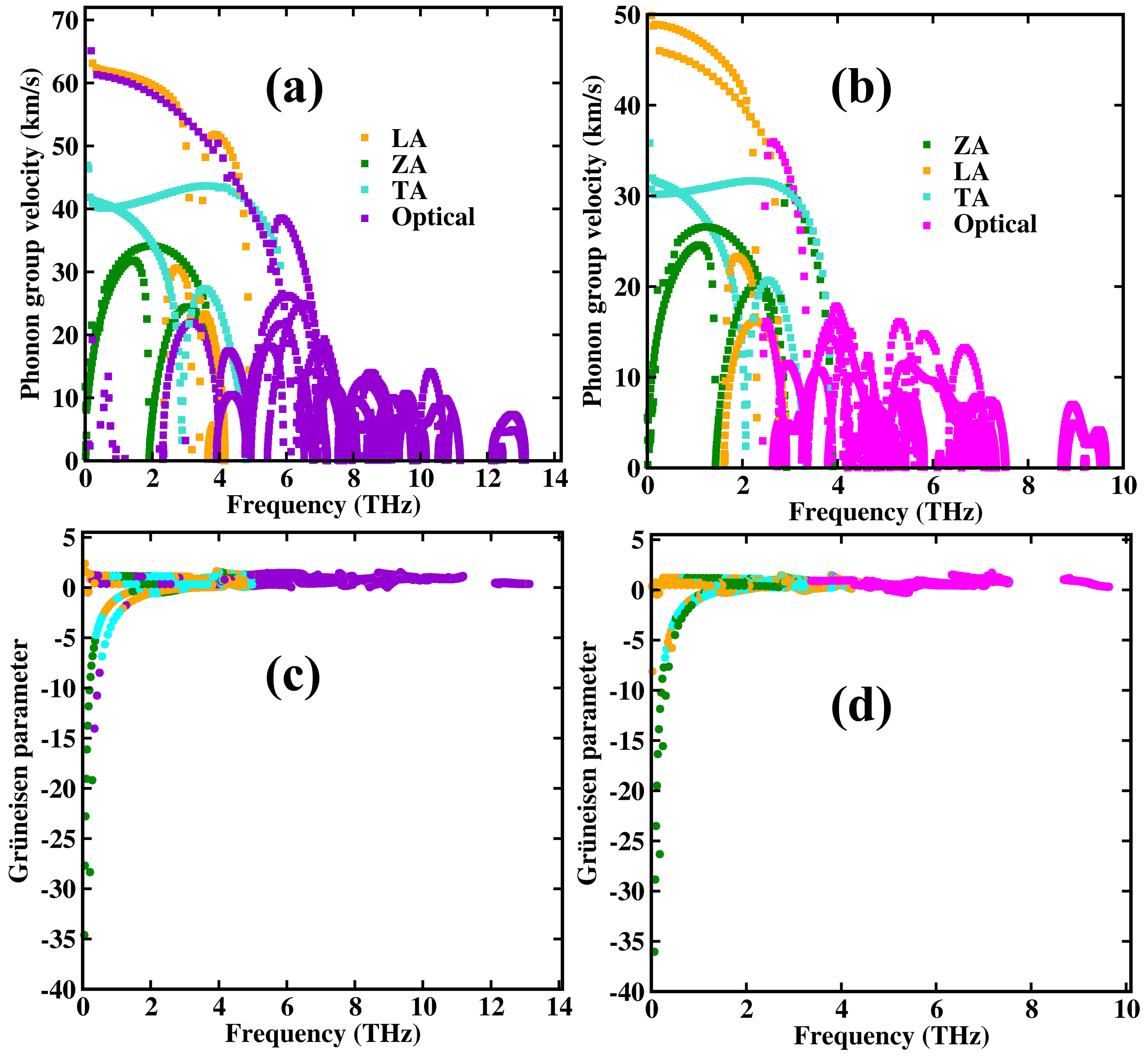}
	\caption{\label{grp_vel_plot} (a) and (b) Shows phonon group velocities and (c) and (d) the Gr\"{u}neisen parameter as a function of phonon frequency for $T^{\prime}-RuS_2$ and $T^{\prime}-RuSe_2$, respectively.}
\end{figure}	

To understand the phonon thermal conductivity, it is important to study phonon group velocity. By calculating these velocities from the slope of the phonon dispersion relations, illustrated in Fig. \ref{grp_vel_plot} (a) and (b), we can assess how phonon dynamics contribute to thermal conductivity. In our calculations, we observe that the LA mode displays the highest phonon group velocities, reaching 65 km/s for $T^{\prime}-RuS_2$ and 50 km/s for $T^{\prime}-RuSe_2$, which are significantly larger than those of the TA and ZA modes. However, the group velocities for optical phonon bands are significantly lower compared to those of acoustic modes. This observation indicates that optical modes predominantly contribute to lowering lattice thermal conductivity, which is beneficial for thermoelectric applications, as it enhances thermoelectric efficiency by reducing the heat carried away by phonons. This variation in phonon group velocities is attributed to the anisotropic bond strength, as the phonon group velocity is directly proportional to the bond strength. The anisotropic nature of the group velocity for both structures is illustrated in Fig. S5(a) and S5(b) of the Supplementary Information.

For optimal thermoelectric materials, minimizing both heat capacity and phonon lifetime is essential. Typically, heat capacity increases with temperature and eventually reaches saturation, as illustrated in Fig. S5(c) and S5(d) of the Supplementary Information. In contrast, phonon lifetime decreases with rising temperature, which contributes to a reduction in $k_{\text{ph}}$. As shown in Fig. S6, phonon lifetime consistently decreases with increasing temperature for both structures. Although both heat capacity and phonon lifetime are factors in determining $k_{\text{ph}}$, the overall reduction in $k_{\text{ph}}$ with temperature suggests that phonon lifetime plays a more significant role as temperature increases.
	
The Gr\"{u}neisen parameter (\(\gamma\)) quantifies the anharmonicity of the phonon spectrum, providing insight into how phonons interact with each other and how lattice vibrations are influenced by changes in volume. A significant aspect of the Gr\"{u}neisen parameter is its relation to thermal conductivity; anharmonic interactions lead to phonon scattering, which affects the mean free path and, consequently, the thermal conductivity. The Gr\"{u}neisen parameter for all phonon branches, including acoustic (ZA, TA, LA) and optical modes illustrated in Fig. \ref{grp_vel_plot} (c) and (d) for both \(T^{\prime}-RuS_2\) and \(T^{\prime}-RuSe_2\). The ZA branch exhibits large negative Gr\"{u}neisen values at low frequencies, indicating significant anharmonicity, while the optical branches remain relatively stable across the frequency range. This stable variation is beneficial for reducing phonon thermal conductivity, as it indicates that anharmonic phonon-phonon scattering remains effective across the vibrational spectrum. Such consistent scattering efficiently reduces phonon lifetimes and mean free paths, leading to suppressed lattice thermal conductivity. This effect is particularly desirable in thermoelectric materials, as it enhances thermal insulation while allowing efficient electronic transport, thereby optimizing the thermoelectric performance of the material.

\subsection{Thermoelectric properties}
An ideal thermoelectric material possesses a seemingly contradictory characteristic: it should be a good conductor of electricity but a poor conductor of heat. Metals excel at conducting electricity but also readily conduct heat, making them unsuitable for thermoelectric applications. Conversely, insulators are excellent thermal insulators but poor electrical conductors. Therefore, semiconductors, with their electrical conductivity lying between metals and insulators, become the prime candidates for thermoelectric studies. They offer a balance between electrical conductivity for efficient current flow and thermal resistance for minimizing heat transfer through the material itself. Our $T^{\prime}-RuX_2$ materials (where $X$= $S$, $Se$) are particularly promising candidates for thermoelectric applications due to their semiconducting nature and favorable bandgaps. $T^{\prime}-RuS_2$ and $T^{\prime}-RuSe_2$ possess indirect bandgaps of $1.69~eV$ and $1.67~eV$, respectively, which is a desirable characteristic for thermoelectrics. Additionally, calculations suggest that these structures exhibit significant phonon thermal conductivity, indicating good heat transport capabilities. This combination of efficient electrical conductivity and moderate thermal conductivity makes the $T^{\prime}-RuX_2$ materials worthy of further investigation for their potential thermoelectric performance.

\begin{table}
		\centering
		\caption{The charge carrier properties including effective mass, elastic modulus, DP constants, mobility, and relaxation time along x- and y-axis calculated at $300~K$ for $T^{\prime}-RuS_2$ and $T^{\prime}-RuSe_2$	materials.}
		\label{tables2}
		\begin{tabular}{ccccccc}
			\hline
			\hline
			& charge & $m^{*}$ & $C_{2D}$ & $E_{DP}$ & $mob_{2D}$ & $\tau$ \\ % <-- added & and content for each column
			& carrier & $(m_0)$ & $(N/m)$ & $(eV)$ & $(cm^{2}/Vs)$ & $(fs)$ \\
			\hline
			$T^{\prime}-RuS_2$ (x) & electron & 0.85 & 150.02 & 15.01 & 21.38 & 10.18 \\ % <--
			$T^{\prime}-RuS_2$ (x) & hole & 0.90 & 150.02 & 14.24 & 19.95 & 10.02 \\
			$T^{\prime}-RuS_2$ (y) & electron & 0.88 & 148.08 & 14.80 & 19.06 & 9.53 \\ % <--
			$T^{\prime}-RuS_2$ (y)& hole & 0.76 & 148.08 & 15.50 & 23.30 & 10.07 \\
			$T^{\prime}-RuSe_2$ (x) & electron & 0.91 & 132.06 & 14.95 & 15.72 & 8.13 \\ % <--
			$T^{\prime}-RuSe_2$ (x) & hole & 0.83 & 132.06 & 15.56 & 18.28 & 8.63 \\ % <--
				$T^{\prime}-RuSe_2$ (y) & electron & 0.76 & 132.30 & 14.15 & 25.16 & 10.87 \\ % <--
			$T^{\prime}-RuSe_2$ (y) & hole & 0.68 & 132.30 & 14.65 & 29.32 & 11.33 \\
			\hline
			\hline	
		\end{tabular}
\end{table}

The electronic Boltzmann transport equations are integrated by the BoltzTraP program to analyze and predict the transport properties of materials under the constant relaxation time and rigid bands approximation. The rigid bands approximation assumes that the electronic band structure of a material remains constant despite doping. In this model, the introduction of dopants shifts the chemical potential (the energy level at which the probability of locating an electron is 50\%) according to the doping concentration and temperature, without affecting the shape or position of the electronic bands. The constant relaxation time approximation (CRTA) assumes that the relaxation time ($\tau$), the average time between scattering events for an electron, does not vary with energy, although it can be temperature-dependent. Under these approximations, the BoltzTraP program calculates transport coefficients, including electrical conductivity and electronic thermal conductivity, as a function of the relaxation time $\tau$.

Our investigation into the thermoelectric properties of $T^{\prime}-RuX_2$ ($X$= $S$, $Se$) materials combines both first-principles calculations and semi-classical Boltzmann transport theory. This approach yields various transport coefficients crucial for thermoelectric evaluation, including the electrical conductivity ($\sigma$/$\tau$), electronic thermal conductivity ($k_{e}/\tau$) and Seebeck coefficient ($S$). As the thermoelectric coefficients generated by BoltzTraP depend on the relaxation time ($\tau$), determining  $\tau$ is necessary for finding the absolute electrical and electronic thermal conductivity of charge carriers. The relaxation time has no impact on the Seebeck coefficients. To the best of our knowledge, no experimental work has been published on calculating the charge carrier relaxation time in $T^{\prime}-RuX_2$.

Theoretical calculations play a crucial role in estimating the relaxation time ($\tau$) in our systems. To this end, we have employed deformation potential (DP) theory, which provides a qualitative understanding of the order of $\tau$, though it does not yield an exact numerical value. The accuracy of electrical conductivity and electronic thermal conductivity under the constant relaxation time approximation (CRTA) is constrained by the undetermined relaxation time, which depends on the specific scattering mechanisms present in the material. According to Matthiessen's rule, \(\tau\) can be expressed as:
\begin{align}
	\frac{1}{\tau} = \frac{1}{\tau_a} + \frac{1}{\tau_{\text{npo}}} + \frac{1}{\tau_{\text{po}}} + \frac{1}{\tau_i} + \frac{1}{\tau_d} + \cdots
\end{align}
where \(\tau_a\), \(\tau_{\text{npo}}\), \(\tau_{\text{po}}\), \(\tau_i\), and \(\tau_d\) correspond to the scattering times due to acoustic phonons, non-polar optical phonons, polar optical phonons, impurities, and defects, respectively. Scattering from impurities and defects is generally temperature-independent and depends on their respective densities. At high temperatures, phonon scattering mechanisms, particularly electron-phonon interactions, become dominant. Electron-phonon scattering is influenced by the electron-phonon coupling matrix, the electronic structure, and the number of phonons, which varies with temperature. Thus, assuming a constant \(\tau\) can introduce significant uncertainties in the calculation of electrical conductivity, power factor, and \(ZT\), particularly when comparing different materials.
	
	For instance, Takagi et al. derived a mobility expression for Si MOSFETs under the assumption that carriers in the inversion layer behave as a 2D electron gas \cite{takagi1994universality}. However, this scenario is far from that of a hypothetical 2D monolayer of a completely different material with a distinct band structure. Such assumptions have led to substantial overestimations of carrier mobility in 2D materials. A notable example is the hole mobility of few-layer black phosphorus, initially estimated to exceed 10,000 cm\(^2\)V\(^{-1}\)s\(^{-1}\) \cite{qiao2014high}, while accurate \textit{ab initio} calculations later provided a much lower estimate of 44 cm\(^2\)V\(^{-1}\)s\(^{-1}\) \cite{sohier2018mobility}. Their review highlights the state-of-the-art methods for mobility calculations in 2D materials and underscores why the DP formalism should be approached with caution. Notably, intervalley phonon scattering plays a crucial role in electron-phonon interactions, and neglecting it in computational studies represents a major limitation. However, despite its limitations, the DP theory remains widely utilized in the study of 2D materials. Many researchers continue to adopt this approach as it offers a computationally efficient and reasonable approximation for transport dominated by acoustic phonon scattering, particularly under the constant electron-phonon coupling approximation in the long-wavelength limit for acoustic phonons \cite{li2021transopt}. While it does not fully account for all scattering mechanisms, it often provides a reasonable order-of-magnitude estimate for relaxation time, as evidenced in several studies \cite{wang2015thermoelectric,kumar2024theoretical,senapati2023thermoelectric}. Therefore, despite its shortcomings, the DP theory remains a tool for estimating transport properties, particularly when computational efficiency is a key consideration.

Looking at Fig. S7 in the Supplementary Information, the VBM and CBM edges in both structures exhibit a parabolic nature, indicating that the energy dispersion near these band edges can be well-approximated by a quadratic equation. This justifies the application of DP theory, which is particularly useful for analyzing how lattice deformations influence band edges and for calculating the effective mass and carrier mobility. Based on this approach, we determined the effective masses for both n-type and p-type carriers in $T^{\prime}-RuS_2$ and $T^{\prime}-RuSe_2$ using the single parabolic band approximation. This model assumes that the energy dispersion near the CBM and VBM follows a quadratic form: $E(k) = E_0 + \frac{\hbar^2 k^2}{2m^*}$, where \(E_0\) is the energy at the band extremum (CBM or VBM), \(\hbar\) is the reduced Planck's constant, \(k\) is the wave vector, and \(m^*\) is the effective mass of the charge carrier. However, if the band structure deviates significantly from a simple quadratic form, higher-order corrections are necessary to ensure accuracy. In such scenarios, the \(\mathbf{k} \cdot \mathbf{p}\) perturbation method provides a more precise evaluation of the effective mass. This method expands the electronic band structure near the band extremum (CBM/VBM) using perturbation theory, incorporating band coupling effects and energy-dependent corrections to the dispersion. The energy dispersion in this approach is given by, $E(k) = E_0 + \frac{\hbar^2 k^2}{2m^*} \left(1 + \eta E(k)\right)$, where \(\eta\) is the non-parabolicity parameter that accounts for deviations from the simple quadratic form. Consequently, the effective mass will be $m^*(E) = m_e (1 + 2\eta E)$ \cite{whalley2019impact}, where $m_e$ is the band-edge effective mass considering parabolic approximation. Moreover, when multiple bands contribute to the conduction and valence bands, the density-of-states (DOS) effective mass (\( m_{\text{DOS}}^* \)) becomes a crucial parameter in determining carrier concentrations in semiconductors. Unlike the single-band effective mass, which is derived from the curvature of an individual band at its extremum, the DOS effective mass accounts for the combined contributions of all participating bands.

\begin{figure}		
	\centering
	\includegraphics[width=1.0\columnwidth]{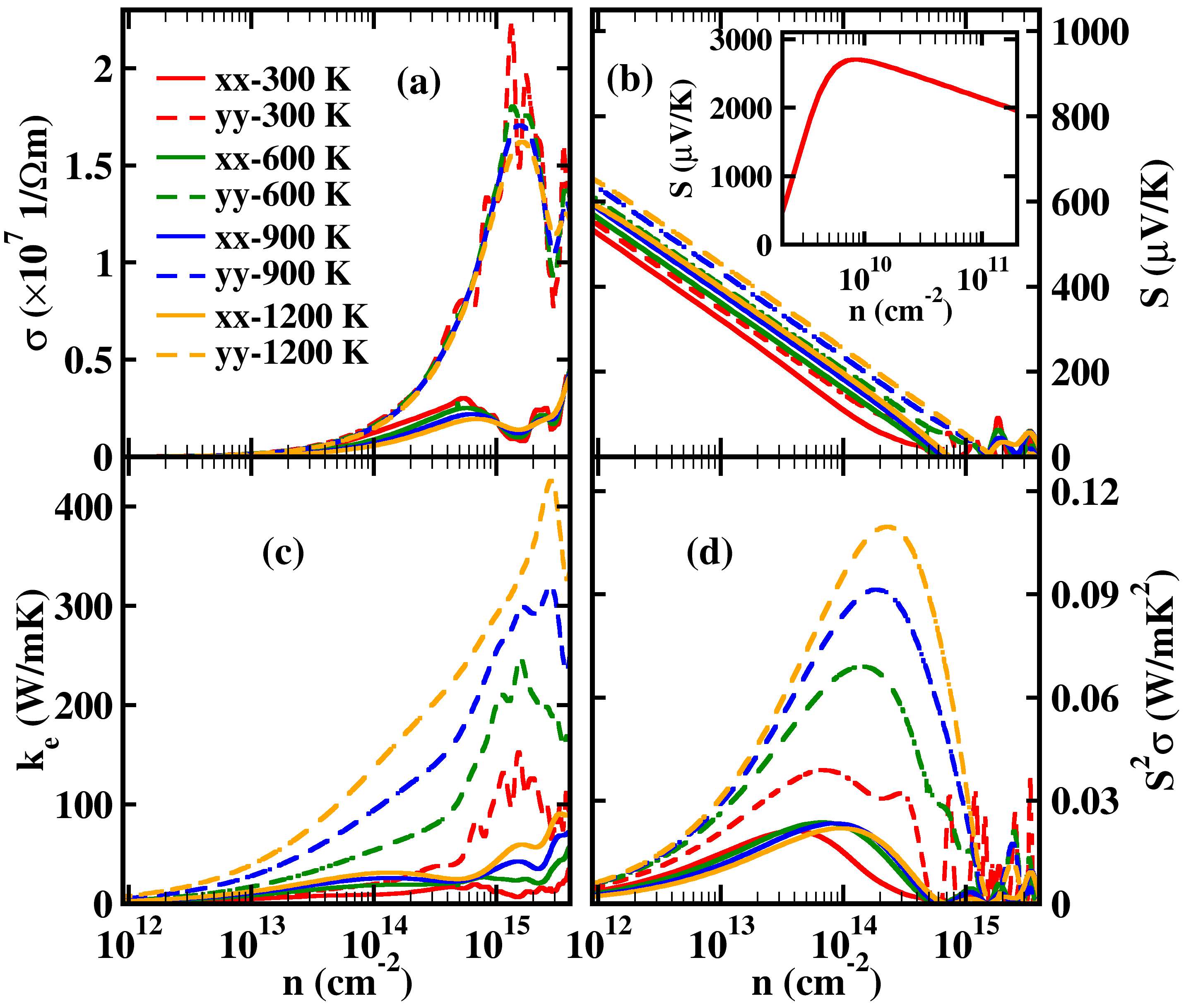}
	\caption{\label{hole_thermo_plot} The variation of (a) electrical conductivity ($\sigma$), (b) Seebeck coefficient ($S$), (c) electronic thermal conductivity ($k_e$) and (d) power factor ($S^{2}\sigma$) as a function of the carrier concentration ($n$) for p-type doped $T^{\prime}-RuS_2$ along x- and y-axis at different temperatures ($T$).}
\end{figure}

Additionally, we calculated the elastic modulus, mobility, relaxation time (at $T = 300~K$), and the deformation potential constant values for both electrons and holes in the materials. These parameters are summarized in Table \ref{tables2}. To obtain the absolute values of the electronic transport coefficients, we employed the average value of $\tau$ (0.101$\times$10$^{-13}$~s) at room temperature. Comparable $\tau$ values are commonly observed in similar 2D materials \cite{senapati2023thermoelectric,wang2015thermoelectric}. This approach ensures that we employ a $\tau$ value that is representative and applicable across the relevant temperature range, thereby maintaining consistency in our calculations and interpretations of electronic transport properties in $T^{\prime}-RuS_2$ and $T^{\prime}-RuSe_2$.

\begin{figure}		
	\centering
	\includegraphics[width=1.0\columnwidth]{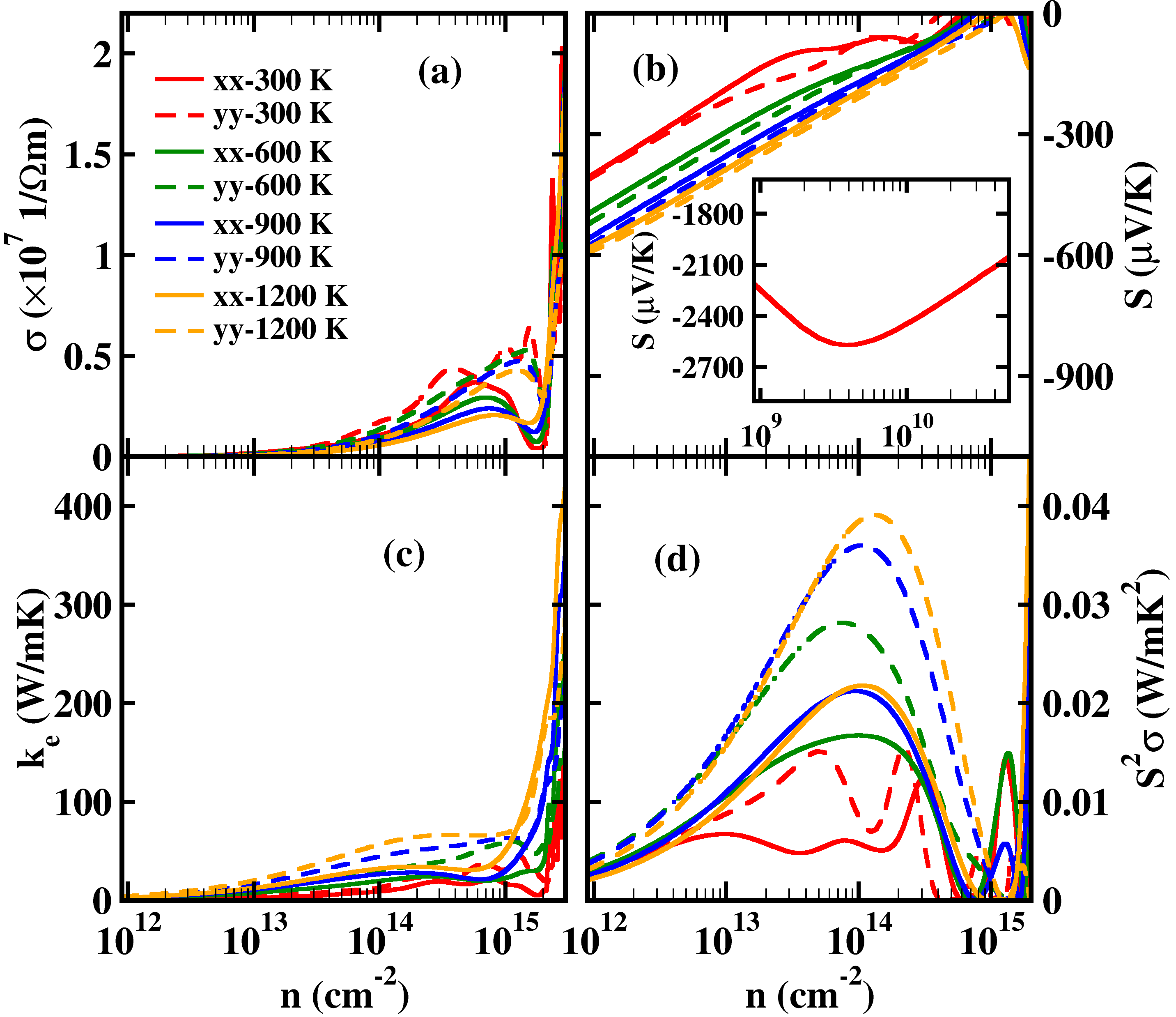}
	\caption{\label{ele_thermo_plot} The variation of (a) electrical conductivity ($\sigma$), (b) Seebeck coefficient ($S$), (c) electronic thermal conductivity ($k_e$) and (d) power factor ($S^{2}\sigma$) as a function of the carrier concentration ($n$) for n-type  doped $T^{\prime}-RuS_2$ along x- and y-axis at different temperatures ($T$).}
\end{figure}

Figure \ref{hole_thermo_plot}(a) illustrates the relationship between anisotropic electrical conductivity ($\sigma$) and carrier concentration ($n$) at various temperatures ($T$) for p-type doped $T^{\prime}-\text{RuS}_2$ along both the x- and y-directions, while Fig. \ref{ele_thermo_plot}(a) shows the same for n-type doping. The observed decrease in electrical conductivity with rising temperature can be attributed to the impact of temperature on electron scattering and carrier mobility. As temperature increases, electrical conductivity decreases due to enhanced electron scattering. Higher temperatures cause more intense lattice vibrations, leading to frequent collisions between charge carriers (electrons or holes) and lattice atoms. This increased scattering reduces carrier mobility and relaxation time, both of which directly impact electrical conductivity. Consequently, as temperature rises, the decrease in carrier mobility results in lower electrical conductivity.

The oscillatory behavior of electrical conductivity as a function of concentration arises from quantum confinement effects and the interaction between charge carriers. As the concentration of charge carriers (electrons or holes) changes, the Fermi level shifts, crossing different electronic subbands that are quantized due to the reduced dimensionality of the material. This leads to variations in the density of states at the Fermi level, which in turn causes the electrical conductivity to oscillate.

Moreover, at specific $T$ and $n$, the $\sigma$ of n-type doped $T^{\prime}-\text{RuS}_2$ is smaller than that of p-type doped $T^{\prime}-\text{RuS}_2$. This difference is attributed to the smaller electron relaxation time compared to that of holes. For instance, at $300~\text{K}$ with a carrier concentration of $10^{15}\text{cm}^{-2}$, the electrical conductivity of n-type doped $T^{\prime}-\text{RuS}_2$ along the y-axis is $5.5\times10^6\Omega^{-1}\text{m}^{-1}$, which is lower than the conductivity of $1.4\times10^7~\Omega^{-1}\text{m}^{-1}$ for the p-type doped system. Along the x-axis, the conductivity values are $3.07\times10^6~\Omega^{-1}\text{m}^{-1}$ for n-type and $2.02\times10^6~\Omega^{-1}\text{m}^{-1}$ for p-type doping. This difference is primarily due to the shorter electron relaxation time compared to that of holes, leading to lower mobility and, consequently, lower electrical conductivity for n-type doping along the y-axis. In contrast, along the x-axis, a slight increase in electron relaxation time relative to holes results in slightly higher conductivity values for n-type doping compared to p-type. Similar trends in electrical conductivity are also observed for $T^{\prime}-\text{RuSe}_2$, and the corresponding plot can be found in Fig. S8 (a) for p-type and Fig. S9 (a) for n-type doped $T^{\prime}-\text{RuSe}_2$ along x- and y-direction in the Supplementary Information.

The thermopower, commonly referred to as the Seebeck coefficient, plays a vital role in assessing thermoelectric performance. It is intrinsically connected to the electronic structure of the material, which means its value is directly influenced by the band structure and the distribution of electronic states. Under the CRTA, the Seebeck coefficient is independent of the relaxation time of charge carriers. This independence arises because, in the formula for $S$, the scattering rate terms cancel out, leaving the Seebeck coefficient solely dependent on the temperature and carrier concentration. Therefore, unlike other transport properties such as electrical conductivity or thermal conductivity, the Seebeck coefficient is fixed by the intrinsic electronic properties of the material without any adjustable parameters related to carrier scattering mechanisms. 

The fluctuation of the Seebeck coefficient ($S$) with carrier concentration ($n$) along x- and y-axis at different temperatures is presented in Fig. \ref{hole_thermo_plot}(b) for p-type and Fig. \ref{ele_thermo_plot}(b) for n-type $T^{\prime}-RuS_2$. The Seebeck coefficient of $T^{\prime}-\text{RuS}_2$ shows a mild anisotropic behavior, with only slight differences observed between the x- and y-directions. As expected, the Seebeck coefficient decreases steadily with increasing carrier density across the entire temperature range, following typical trends. This reduction in the Seebeck coefficient with higher carrier concentration aligns with the general behavior in thermoelectric materials, where increased carrier density enhances the electrical conductivity but reduces the thermoelectric voltage generated per unit temperature difference. We find $T^{\prime}-RuS_2$ display a high $S$ (larger than $2500~\mu V/K$ from the inset of Fig. \ref{hole_thermo_plot} (b) and Fig. \ref{ele_thermo_plot} (b)) within a reasonable substantial range of carrier concentrations ($10^{10}~cm^{-2}$), regardless of p-type or n-type. Due to degenerate states in valence band compared to conduction band, p-type $S$ is somewhat superior to that of n-type. Specifically, at $300~K$, the maximum Seebeck coefficients for $T^{\prime}-RuS_2$ are $2685~\mu V/K$ for p-type and $2585~\mu V/K$ for n-type (see the inset of Fig. \ref{hole_thermo_plot} (b) and Fig. \ref{ele_thermo_plot} (b)). The higher Seebeck coefficients observed in $T^{\prime}-RuS_2$ compared to many TMDs are advantageous for enhancing the power factor in thermoelectric applications \cite{purwitasari2022high,bera2022low,bera2019strain}.

The plots of Seebeck coefficients for $T^{\prime}-RuSe_2$ along x- and y-axis can be found in Fig. S8(b) for p-type and Fig. S9(b) for n-type in the supplementary information. In contrast to $T^{\prime}-RuS_2$, we observe slightly different trends for $T^{\prime}-RuSe_2$. The values of $S$ in $T^{\prime}-RuSe_2$ are lower compared to those in $T^{\prime}-RuS_2$. These differences can be attributed to variations in the electronic band structure and carrier dynamics between $T^{\prime}-RuS_2$ and $T^{\prime}-RuSe_2$.

Figure \ref{hole_thermo_plot}(c) shows the effect of carrier concentration at various temperatures on the electronic thermal conductivity ($k_e$) along the x- and y-axes for p-type $T^{\prime}-\text{RuS}_2$, while Fig. \ref{ele_thermo_plot}(c) illustrates this effect for n-type $T^{\prime}-\text{RuS}_2$. In this context, we observed that $k_e$ rises with increasing temperature and carrier concentration. This increase can be attributed to the larger number of charge carriers that contribute to thermal transport. As the temperature increases, more charge carriers gain enough energy to overcome the bandgap, leading to a significant rise in the availability of free electrons or holes for heat conduction. This enhanced population of carriers directly enhances $k_e$ by allowing more particles to effectively transfer thermal energy through the material. The highest optimal $k_e$ value for p-type $T^{\prime}-RuS_2$ is significantly larger ($143~W/mK$) at $300~K$ compared to n-type $T^{\prime}-RuS_2$ ($35.82~W/mK$). This difference can be attributed to the fact that p-type $T^{\prime}-RuS_2$ has lighter effective masses for holes compared to n-type $T^{\prime}-RuS_2$ for electrons. Lighter effective mass enables holes to move more freely through the material, facilitating more efficient thermal transport and resulting in higher $k_e$ values. Therefore, the lighter effective mass of holes in p-type $T^{\prime}-RuS_2$ enhances its ability to conduct heat compared to the heavier electrons in n-type $T^{\prime}-RuS_2$. Moreover, the influence of carrier concentration at various temperatures on $k_e$ along x- and y-axis are depicted in Fig. S8(c) for p-type and Fig. S9(c) for n-type $T^{\prime}-RuSe_2$ in the supplementary information. 

According to the Wiedemann-Franz law, \( k_e = L \sigma T \), where \( L \) is the Lorentz number, \( \sigma \) is the electrical conductivity, and \( T \) is the temperature. To validate \( L \), we calculated its value at 300~K, 600~K, 900~K, and 1200~K across a range of concentrations, from low to high, for both n-type and p-type T\('\)-RuS\(_2\) structures. The results are reported in Table~SI of the supplementary information which is also consistent with the literature \cite{ghosh2020thermal,thesberg2017lorenz}. From the table, we observe that moderate-to-high concentration ranges effectively balance electrical and thermal conductivity, confirming that the Wiedemann-Franz law holds true within this concentration range as temperature increases.

\begin{figure}		
	\centering
	\includegraphics[width=0.9\columnwidth]{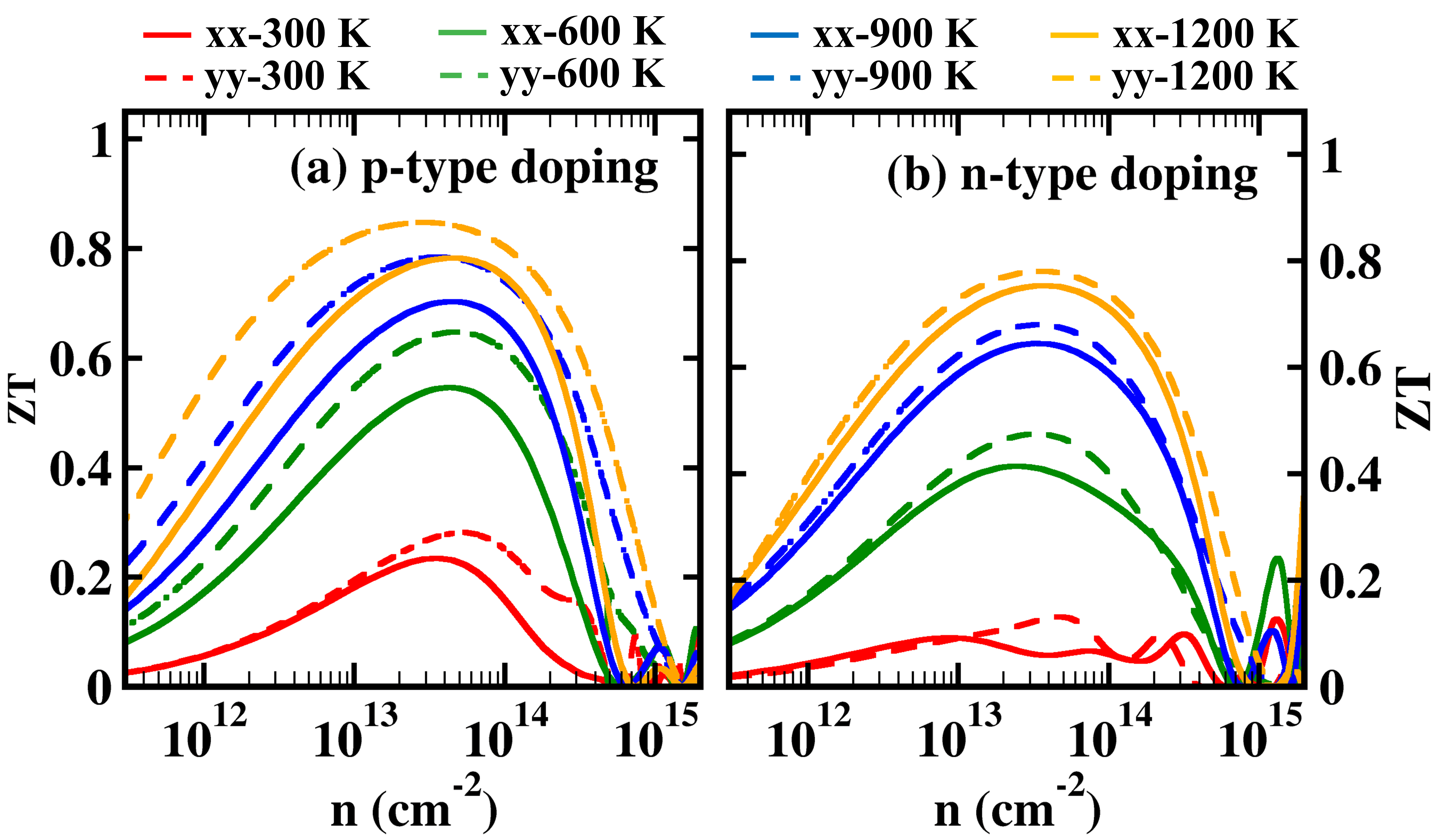}
	\caption{\label{zt_plot} The variation of thermoelectric figure of merit ($ZT$) as a function of the carrier concentration ($n$) for (a) p-type and (b) n-type doped $T^{\prime}-RuS_2$ along x- and y-axis at different temperatures ($T$).}
\end{figure}

The power factor (PF) is a crucial parameter for assessing the electronic transport properties of a material. It measures the interplay between the electrical conductivity and Seebeck coefficient, highlighting their combined effects on thermoelectric performance. Specifically, $\sigma$ tends to increase with rising carrier concentration, whereas $S$ is inversely proportional to carrier concentration. Thus, achieving an optimal power factor requires balancing the carrier concentration to ensure that both $S$ and $\sigma$ are maximized. This balance ensures that both $S$ and $\sigma$ contribute effectively to enhancing the overall thermoelectric performance of the material. The calculated PF of $T^{\prime}-RuS_2$ for p-type and n-type systems along x- and y-axis are plotted in Fig. \ref{hole_thermo_plot}(d) and Fig. \ref{ele_thermo_plot}(d), respectively. The optimal PF values for p-type systems are higher than those for n-type systems, primarily due to the larger electrical conductivity of the p-type materials. As the temperature increases, an increment in PF is observed. Notably, the highest PF value for p-type $T^{\prime}-RuS_2$ is $0.03~W/mK^{2}$ along y-axis at $300~K$, which is three times larger than the $0.01~W/mK^{2}$ for n-type $T^{\prime}-RuS_2$.
This optimal PF value for p-type $T^{\prime}-RuS_2$ is also significantly higher than those of many TMDs thermoelectric materials at room temperature, such as $MoS_2$ ($\approx0.0085~W/mK^{2}$) \cite{hippalgaonkar2017high} and $WSe_2$ ($\approx0.01~W/mK^{2}$)\cite{kumar2015thermoelectric}. Furthermore, at $1200~K$, the highest power factor for p-type $T^{\prime}-RuS_2$ can reach up to $0.10~W/mK^{2}$ along y-axis, while for n-type $T^{\prime}-RuS_2$ it can reach up to $0.038~W/mK^{2}$.
Similar trends are also found for both p-type and n-type $T^{\prime}-RuSe_2$, with the PF being less than that of $T^{\prime}-RuS_2$. The plots for PF for both p-type and n-type $T^{\prime}-RuSe_2$ along x- and y-axis can be found in the supplementary information, in Fig. S8(d) and Fig. S9(d), respectively.

The figure of merit, $ZT$, exhibits variations with carrier concentration for $T^{\prime}-RuS_2$ at different temperatures along x- and y-direction, as depicted in Fig. \ref{zt_plot}(a) for p-type and Fig. \ref{zt_plot}(b) for n-type doping. Our findings underscore the significant impact of doping type on the $ZT$ values of $T^{\prime}-RuS_2$ at specific temperatures. Specifically, at a constant carrier concentration of $10^{13}~cm^{-2}$ and within the temperature range of $300$ to $1200~K$,  the maximum $ZT$ values for p-type doped $T^{\prime}-\text{RuS}_2$ are $0.78$ along the x-axis and $0.85$ along the y-axis. For n-type doped $T^{\prime}-\text{RuS}_2$, the maximum $ZT$ values reach $0.75$ along the x-axis and $0.78$ along the y-axis. This difference is primarily attributed to the higher electrical conductivity of holes compared to electrons in p-type doping, thereby enhancing the thermoelectric performance as indicated by the $ZT$ values. These $ZT$ values are also comparable to many promising thermoelectric TMDs \cite{tao2020thermoelectric,pallecchi2020review}.

As the temperature increases from $300~\text{K}$ to $1200~\text{K}$, both p-type and n-type $ZT$ values also increase. Specifically, the maximum $ZT$ values rise from $0.27$ to $0.85$ for p-type doping and from $0.13$ to $0.78$ for n-type doping along y-direction. This trend suggests that p-doped $T^{\prime}-\text{RuS}_2$ along y-axis shows promise as a favorable thermoelectric material for applications around $1200~\text{K}$. The higher $ZT$ values observed at high temperatures are predominantly contributed by the lower phonon thermal conductivity ($k_{ph}$) in the material. Moreover, for $T^{\prime}-\text{RuSe}_2$, at a temperature of $1200~\text{K}$ and a constant carrier concentration of $10^{13}~\text{cm}^{-2}$, the $ZT$ values reach $0.81$ for p-type and $0.69$ for n-type doping along the x-axis, and $0.87$ for p-type and $0.78$ for n-type doping along the y-axis. These values can be found in Fig. S10(a) for p-type and Fig. S10(b) for n-type doping in the supplementary information. These $ZT$ values for both $T^{\prime}-\text{RuS}_2$ and $T^{\prime}-\text{RuSe}_2$ highlight their potential for efficient thermoelectric energy conversion in high-temperature environments.

\section{Conclusions}
In summary, we conducted first-principles calculations to investigate the structural stability and thermoelectric performance of distorted TMDs $T^{\prime}-\text{RuS}_2$ and $T^{\prime}-\text{RuSe}_2$. Our analysis of cohesive energy, formation energy, phonon dispersion spectra, and elastic constants indicates that these structures are energetically and mechanically stable, making them favorable candidates for laboratory synthesis.

Thermoelectric transport properties of both structures were studied using a combination of first-principles calculations and semi-classical Boltzmann transport theory. Our study highlights the significant thermoelectric potential of $T^{\prime}-\text{RuS}_2$ and $T^{\prime}-\text{RuSe}_2$. At $300~\text{K}$, $T^{\prime}-\text{RuS}_2$ demonstrates high Seebeck coefficients of $2685~\mu\text{V/K}$ for hole doping and $2585~\mu\text{V/K}$ for electron doping. In contrast, $T^{\prime}-\text{RuSe}_2$ exhibits Seebeck coefficients of $1515~\mu\text{V/K}$ for hole doping and $1533~\mu\text{V/K}$ for electron doping. These high Seebeck coefficients highlight their suitability for thermoelectric applications.

Furthermore, both materials demonstrate favorable power factors, with notable $ZT$ values. For p-type $T^{\prime}-\text{RuS}_2$ and $T^{\prime}-\text{RuSe}_2$, the maximum $ZT$ values reach $0.85$ and $0.87$, respectively, at $1200~\text{K}$, while for n-type doping, the values are $0.78$ for both materials along y-axis. Here, p-type $T^{\prime}-\text{RuS}_2$ and $T^{\prime}-\text{RuSe}_2$ exhibit superior thermoelectric performance compared to their n-type counterparts, attributed to higher electrical conductivity and lower phonon thermal conductivity. Thus, $T^{\prime}-\text{RuS}_2$ and $T^{\prime}-\text{RuSe}_2$ show promise as efficient thermoelectric materials, particularly suited for high-temperature applications. To further enhance their thermoelectric performance and achieve higher ZT values, several strategies can be explored, including optimization through strain engineering, nanostructuring, and dimensional reduction \cite{ruan2022strain,zhang2021potential}. This study contributes to advancing our understanding of TMDs and their potential role in developing advanced thermoelectric devices.

\section*{Acknowledgement}
PS acknowledges financial support from DST-INSPIRE (IF190005). AK thanks the University Grants Commission (UGC), New Delhi, for providing financial assistance through a Senior Research Fellowship (DEC18-512569-ACTIVE). PP acknowledges support from DST-SERB for the ECRA project (ECR/2017/003305).

\section*{Author contributions}
Parbati Senapati contributed to conceptualization, methodology, formal analysis, investigation, writing of the original draft; Ajay Kumar contributed to formal analysis, methodology, reviewing, and editing of the manuscript; Prakash Parida contributed to supervision, reviewing, and editing of the manuscript.

\section*{Data availability}
The datasets are available from the corresponding author on reasonable request.
\section*{Declarations}	
There is no conflict of interest.

\section*{Supplemetary data}
The supplementary material includes detailed convergence tests, covering the k-mesh grid and plane wave energy cutoff for electronic properties, as well as the supercell convergence test for phonon dispersion. Additionally, it presents the q-grid convergence test and cutoff for the third-order force constant in phonon thermal conductivity calculations. The material also provides phonon group velocity data along the x- and y-directions, phonon lifetime as a function of frequency, and the variation of key thermoelectric parameters—including electrical conductivity, thermopower, power factor, electronic thermal conductivity, and the figure of merit—along both axes for n-type and p-type $T^{\prime}-RuSe_2$.

%\bibliography{bibliography.bib}
%\bibliographystyle{siam}

\end{document}